\documentclass{article}
\usepackage{pdfpages}
\usepackage{enumitem}
\usepackage{graphicx}
\usepackage{wrapfig}
\usepackage{multicol} 
\usepackage{authblk}  
\usepackage{url}
\usepackage{natbib}
\usepackage{fullpage}
\usepackage{hyperref}
\usepackage{tabularx}
\usepackage{amsmath}
\usepackage{float} 

\title{Safety Cases: How to Justify the Safety of Advanced AI Systems}
\author{Joshua Clymer\thanks{Correspondence to joshuamclymer@gmail.com. Author contributions are described in Appendix \ref{sec:author_contributions}.}}
\author[2]{Nicholas Gabrieli}
\author[3]{David Krueger}
\author[4]{Thomas Larsen}

\affil[1]{Columbia University}
\affil[2]{Harvard University}
\affil[3]{University of Cambridge}
\affil[4]{Center for AI Policy}

\date{March 6, 2024} 

\begin{document}
\includepdf{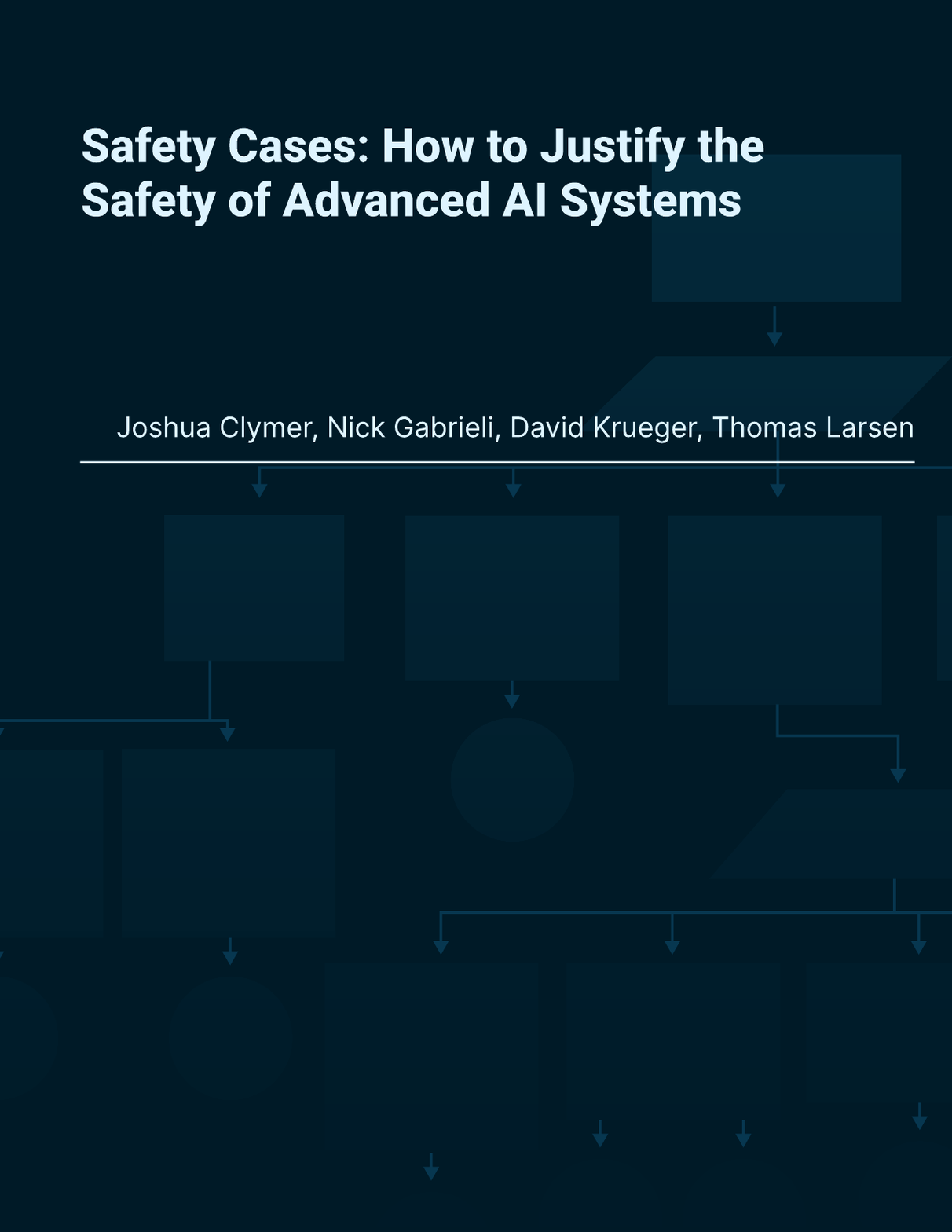}

\maketitle
\begin{center}
\begin{minipage}{0.8\textwidth}
\begin{center}
    \fontsize{12pt}{14pt}\textbf{Abstract}
\end{center}
\noindent\fontsize{12pt}{14pt}\selectfont
As AI systems become more advanced, companies and regulators will make difficult decisions about whether it is safe to train and deploy them. To prepare for these decisions, we investigate how developers could make a ‘safety case,’ which is a structured rationale that AI systems are unlikely to cause a catastrophe. We propose a framework for organizing a safety case and discuss four categories of arguments to justify safety: total \textbf{inability} to cause a catastrophe, sufficiently strong \textbf{control} measures, \textbf{trustworthiness} despite capability to cause harm, and -- if AI systems become much more powerful -- \textbf{deference} to credible AI advisors. We evaluate concrete examples of arguments in each category and outline how arguments could be combined to justify that AI systems are safe to deploy.
\end{minipage}
\end{center}

\newpage

\begin{center}
    \includegraphics[width=1\textwidth]{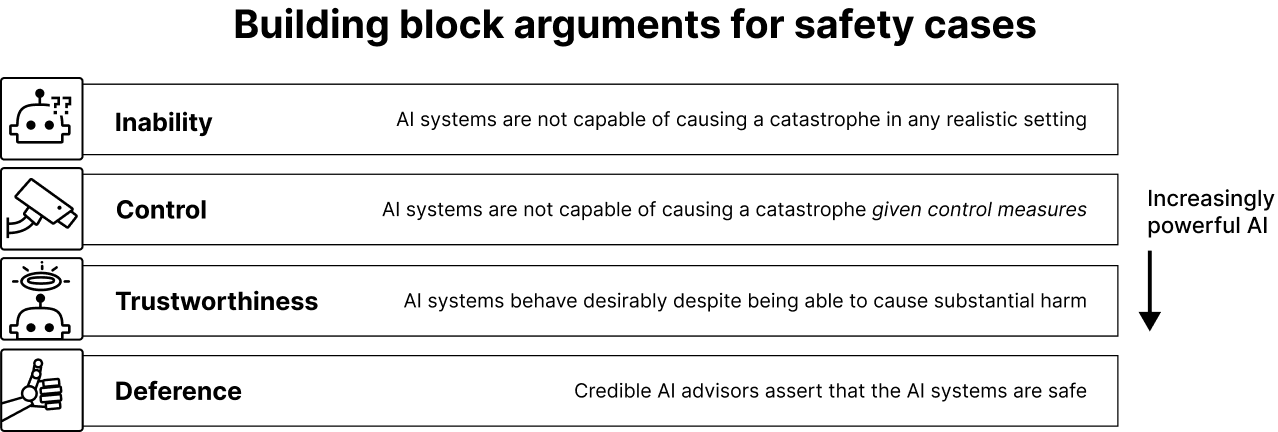}
\end{center}

\tableofcontents
\newpage

\section[Introduction]{\LARGE Introduction}
\label{introduction}

AI experts have recently called attention to the risks posed by advanced AI systems. For example, Geoffrey Hinton, who has been called a ‘godfather’ of modern AI, recently left Google to raise the alarm about future risks from AI systems \citep{MIT_tech_hinton}. AI could cause many societal harms \citep{toreini2022technologies}. The most extreme dangers that Hinton and others have warned about include the risk that AI systems are \textbf{misused} or \textbf{escape human control} \citep{hendrycks2023overview}.

To mitigate these dangers, researchers have called on developers to provide evidence that their systems are safe \citep{koessler2023risk, schuett2023best}; however, \textbf{the details of what this evidence should look like have not been spelled out}. For example, Anderljung et al vaguely state that this evidence should be ``informed by evaluations of dangerous capabilities and controllability"\citep{anderljung2023frontier}. Similarly, a recently proposed California bill asserts that developers should provide a ``positive safety determination" that “excludes hazardous capabilities” \citep{CaliforniaSB1047}. These nebulous requirements raise questions: what are the core assumptions behind these evaluations? How might developers integrate other kinds of evidence? This report aims to answer these questions by providing a detailed overview of \textit{how developers could justify that their AI systems are safe to deploy.}

We first introduce the concept of a “safety case,” which is a method of presenting safety evidence used in six industries in the UK \citep{SUJAN2016181}. A safety case is a structured rationale that a system is unlikely to cause significant harm if it is deployed to a particular setting. The remainder of the report then explains how developers could construct a safety case in the context of catastrophic AI risk.

Section \ref{sec:executive} provides an executive summary.

Section \ref{sec:definitions} defines essential terminology.

Section \ref{sec:recommendations} lists recommendations for institutions that use safety cases to make AI deployment decisions.

Section \ref{sec:arguments} groups arguments that AI systems are safe into four categories: inability, control, trustworthiness, and deference. It then describes examples of arguments in each category and rates their practicality, maximum strength, and scalability.

Finally, section \ref{sec:holistic} explains how to combine the arguments from section \ref{sec:arguments} into a holistic safety case.

\newpage

\section[Executive Summary]{\LARGE Executive Summary}
\label{sec:executive}

\includegraphics[width=1\textwidth]{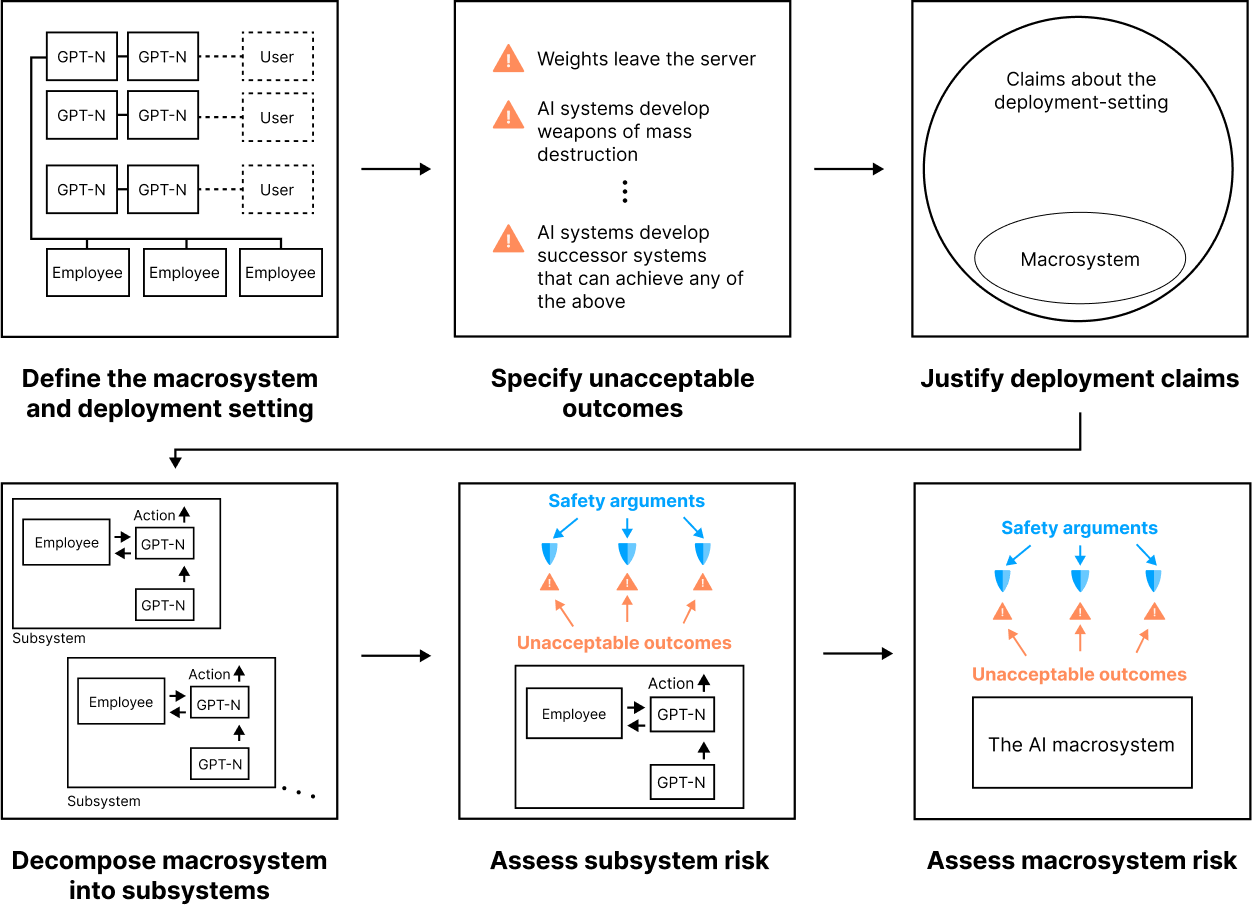}

\subsection{A framework for structuring an AI safety case}

We propose a framework for structuring a safety case that is modeled from traditional safety analysis frameworks like Failure Modes and Effects Analysis (FMEA) \citep{FEMA}. It is composed of six steps:
\begin{enumerate}[label=\arabic*.,leftmargin=*,labelindent=\parindent, noitemsep,topsep=0pt,parsep=0pt,partopsep=0pt]

\item \textbf{Define the AI macrosystem and deployment decision:}  The AI macrosystem is the collection of all AI systems and supporting infrastructure that developers wish to deploy. Evaluators of a safety case must first understand what developers are proposing to do: how is the macrosystem configured and what setting would it operate in?
\item \textbf{Specify unacceptable outcomes:} Developers then decompose the somewhat abstract claim that the AI macrosystem “won’t cause a catastrophe” into more concrete threat models such as “AI system weights won’t leave the server,” “AI systems will not develop weapons of mass destruction,” etc. The remainder of the safety case argues that these outcomes will not occur.
\item \textbf{Justify deployment assumptions:} Next, developers lay additional groundwork by justifying claims about the deployment setting; for example, “AI system weights are secure from human actors” or “companies with fine-tuning access will follow the terms of service.” Once claims about the deployment setting are established, developers turn to how the behavior of their AI systems could cause a catastrophe.
\item \textbf{Decompose the macrosystem into subsystems:} Modern AI macrosystems are large (millions of models) and interact with many different services and environments. In order to make the macrosystem easier to analyze, developers could first analyze the behavior of smaller subsystems. We’ll call these subsystems \textbf{subsystems}. An example of a subsystem is a GPT agent combined with a classifier that monitors its outputs.
\item \textbf{Assess subsystem risk:} Developers first identify ways that each subsystem could \textbf{unilaterally} achieve an unacceptable outcome (i.e. without the meaningful involvement of other subsystems). Then, they must argue that each subsystem has an acceptably low probability of doing so.
\item \textbf{Assess macrosystem risk:} Finally, developers identify ways \textbf{interactions} between subsystems could cause an unacceptable outcome and argue these threat models are also unlikely.

\end{enumerate}

Following UK standards, we summarize safety cases in goal structuring notation (GSN), which is a flow-chart diagram that illustrates how claims are connected \citep{GSN_standard}. Optionally, probabilities are assigned to each node and aggregated to yield an overall risk estimate \citep{space_shuttle}. We explain the end-to-end process of using our framework in section \ref{sec:holistic}.

\begin{figure}[ht]
\centering
\includegraphics[width=1\textwidth]{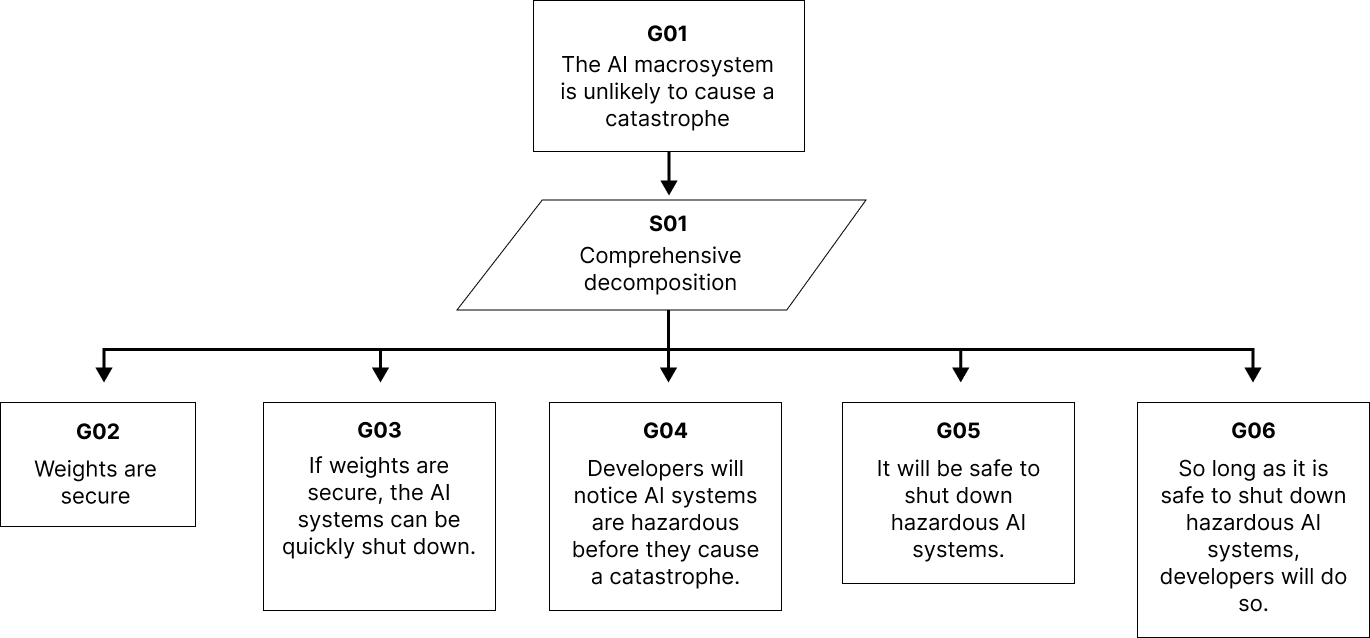}
\caption{the GSN diagram is the start of a holistic safety case \ref{sec:holistic}. The decomposition above would occur in step 2 (specifying unacceptable outcomes). ‘G’ labeled rectangles represent subclaims (i.e. goals). ‘S’ labeled parallelograms indicate justification strategies. For an example of a full safety case, see section \ref{sec:holistic}.}
\label{fig:GSN}
\end{figure}

\pagebreak
\subsection{Safety argument categories}

Developers could use a wide variety of arguments to justify the safety of the subsystems of a macrosystem \ref{sec:arguments}. To provide a clear picture of what these arguments could look like, we describe four categories: inability, control, trustworthiness, and deference.\\~\\

\noindent\begin{minipage}{\textwidth}
  \raisebox{-0.55\height}{\includegraphics[width=0.075\textwidth]{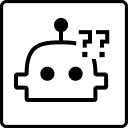}}
  \hfill 
  \begin{minipage}[t]{0.91\textwidth}
    \textbf{\large Inability arguments: “AI systems are incapable of causing unacceptable outcomes in any realistic setting.”}
  \end{minipage}
\end{minipage}

Inability arguments establish that AI systems could not cause a catastrophe in any setting they could realistically be in; for example, even if they were placed on unmonitored servers and given access to substantial computation resources, they would still be unable to achieve unacceptable outcomes. 

The standard structure of inability arguments is as follows:
\begin{itemize}[noitemsep,topsep=0pt,parsep=0pt,partopsep=0pt]
\item A red team has identified strategies that users and/or AI systems could pursue to cause unacceptable outcomes if the AI systems were unrestricted (e.g. open-sourced).
\item The red team has designed `proxy tasks' intended to be such that if AI systems are unable to accomplish them, then the AI systems are unable to execute the strategies identified in the first step.
\item The AI systems are unable to accomplish the proxy tasks even after being fine-tuned\footnote{Currently, the gold standard fine-tuning strategy for making inability arguments is supervised fine-tuning. If supervised fine-tuning does not succeed in eliciting the capability, then this is strong evidence that an AI system is not capable of accomplishing a proxy task – at least by pursuing strategies represented in the supervised fine-tuning data. Ideally, other approaches like RL are also used; however, these methods don’t provide the same level of negative evidence if they fail to be effective.} to do so (for example, using human demonstration data).
\item Fine-tuning reliably elicits the capabilities of the AI systems to perform the proxy tasks.
\item The AI systems will continue to be incapable of accomplishing the proxy tasks even as they learn or are modified during the deployment window.
\item The red team is sufficiently competitive with users and the AI systems at identifying strategies for causing unacceptable outcomes.
\end{itemize}

The last point could be difficult to justify because the red team is unlikely to consider all strategies for causing unacceptable outcomes. For example, one broad strategy users might employ is to integrate them with other software called ‘scaffolding.’ Red teams will likely not discover optimal scaffolding for any given task. So, they must argue that they are \textit{sufficiently competitive} relative to bad actors. For example, they could argue that scaffolding advances will be small enough that they are unlikely to make AI systems dangerous, or that progress can be monitored and reassessments can be made in light of scaffolding advances.\\~\\

\noindent\begin{minipage}{\textwidth}
  \raisebox{-0.55\height}{\includegraphics[width=0.075\textwidth]{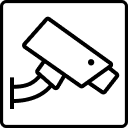}}
  \hfill 
  \begin{minipage}[t]{0.91\textwidth}
    \textbf{\large Control arguments: “AI systems are incapable of causing unacceptable outcomes given existing control measures.”}
  \end{minipage}
\end{minipage}

Control arguments are similar to inability arguments as both depend on the capabilities of AI systems; however, control arguments additionally depend on environmental measures that prevent AI systems from causing unacceptable outcomes. For example, a control argument might show that, even though the AI systems could develop a weapon of mass destruction if they had unrestricted access to the internet, monitoring systems prevent them from doing so.

Control arguments have the following standard structure:
\begin{itemize}[noitemsep,topsep=0pt,parsep=0pt,partopsep=0pt]

\item Control measures impose some set of restrictions on AI systems.
\item Users and AI systems are incapable of evading control measures to cause unacceptable outcomes (applying the same structure of argument from inability).

\end{itemize}

Control arguments conservatively assume that some subset of models and users pursue their \textit{best available strategy} to cause a catastrophe. A red team then attempts to enumerate these strategies and show that users and models can’t successfully execute them in the presence of existing control measures.\\~\\

\noindent\begin{minipage}{\textwidth}
  \raisebox{-0.75\height}{\includegraphics[width=0.075\textwidth]{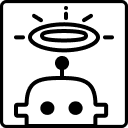}}
  \hfill 
  \begin{minipage}[t]{0.91\textwidth}
    \textbf{\large Trustworthiness arguments: “Even if AI systems are capable of causing catastrophic outcomes, they will not do so because they robustly behave in an intended way.”}
  \end{minipage}
\end{minipage}

Trustworthiness arguments claim that AI systems are safe because they consistently behave in a desirable way. For example, trustworthy behavior arguments could assert that AI systems are are robust to jailbreaks, are truthful, follow developer-specified rules, etc.

Trustworthiness arguments have the following standard structure:
\begin{itemize}[noitemsep,topsep=0pt,parsep=0pt,partopsep=0pt]

\item The behavior of the AI systems has been reliably evaluated in a wide variety of circumstances. Wherever evaluations are reliable, the behavior of the AI systems is consistent with a behavioral property that implies the AI systems would not achieve an unacceptable outcome. For example, “the AI systems follow developer-specified rules.”
\item The AI systems will continue to exhibit this behavioral property during the deployment window.
\end{itemize}

It could be challenging to justify that AI systems will continue behaving as intended during deployment because of \textbf{distribution shift} and \textbf{black swans}. 

\textbf{Distribution shift} occurs when the deployment environment is different from the environment where the AI systems are evaluated. For example, a distribution shift would occur if evaluators expend less effort to find jailbreaks than users collectively expend during deployment (section \ref{sec:distribution_shift}).

\textbf{Black swans} are rare inputs that cause AI systems to catastrophically misbehave. Black swans may be rare enough to where they are not sampled during evaluation but are common enough to occur in deployment (section \ref{sec:blackswans}).

We discuss two important argument templates developers could use to justify that black swans and distribution shift won’t cause AI systems to behave badly: 
\begin{enumerate}
    \item \textbf{Not alignment faking arguments}: These arguments claim that AI systems are unlikely to cause a catastrophe unless they are faking alignment. By `faking alignment,' we mean that the AI systems are deliberately waiting for a good opportunity to misbehave. These arguments then show that AI systems are unlikely to be faking alignment. (section \ref{subsubsec:naf}).
    \item \textbf{Eliciting latent knowledge arguments}: These arguments claim that AI systems would ‘know’ (i.e. internally represent) whether their actions are safe. They then claim that AI systems believe their actions are harmless (section \ref{subsubsec:ELK}).
\end{enumerate}

Justifying the safety of advanced AI systems with “not alignment faking” or “eliciting latent knowledge” arguments will likely require fundamental research advances (Section 5.4).\\~\\

\noindent\begin{minipage}{\textwidth}
  \raisebox{-0.75\height}{\includegraphics[width=0.075\textwidth]{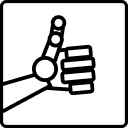}}
  \hfill 
  \begin{minipage}[t]{0.91\textwidth}
    \textbf{\large Deference arguments: “AI advisors assert that other AI systems do not pose a catastrophic risk, and these AI advisors are at least as credible as human decision-makers.”}
  \end{minipage}
\end{minipage}

If AI systems are deemed credible with a trustworthiness argument, they could be consulted about whether other AI systems are safe. Evaluators could then use the claims of AI advisors as evidence similar to how regulators often rely on the judgments of human experts as evidence \citep{NRC-experts}.

Deference arguments have the following standard structure:
\begin{itemize}[noitemsep,topsep=0pt,parsep=0pt,partopsep=0pt]

\item AI advisors are capable of providing judgments about the safety of AI systems that are at least as reliable as those of human decision-makers.
\item AI advisors are trustworthy in the sense that they consistently provide these high-quality judgments.

\end{itemize}

As AI systems become increasingly advanced, developers will likely rely on argument categories in the progression in which they have been presented. First, developers will mostly rely on inability, then control, then trustworthiness. Eventually, AI systems and their deployment environments may become sufficiently complex to where humans must justify their safety by deferring to AI advisors.

\begin{figure}[ht]
\centering
\includegraphics[width=.6\textwidth]{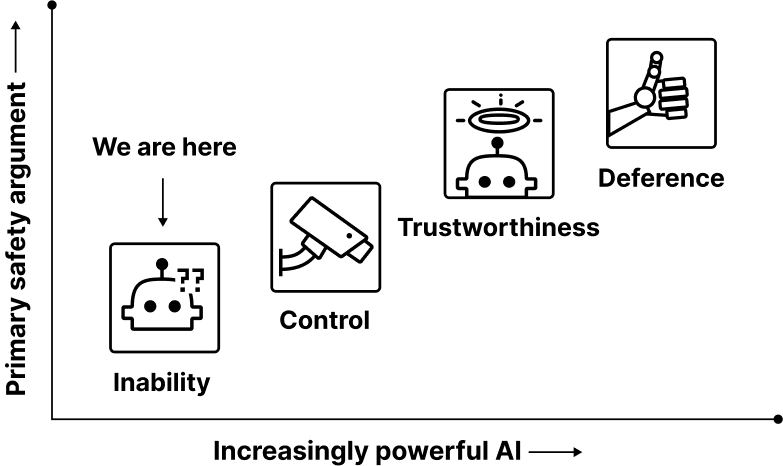}
\caption{As AI systems become more powerful, developers will likely increasingly rely on arguments toward the right in the plot above.}

\label{fig:args_over_time}
\end{figure}
\pagebreak
\subsection{Safety argument examples}

In section \ref{sec:arguments}, we list \textbf{sixteen examples of arguments} that could be included in a holistic safety case.

We rank each argument on the following dimensions:

\begin{itemize}[noitemsep,topsep=0pt,parsep=0pt,partopsep=0pt]
\item \textbf{Practicality:} Could the argument be made soon or does it require substantial research progress?
\item \textbf{Maximum strength:} How much confidence could the argument give evaluators that the AI systems are safe?
\item \textbf{Scalability:} To what extent can the argument be applied to very powerful AI systems?
\end{itemize}

We provide a more detailed explanation of these labels in section \ref{subsec:desiderata}.

\begin{figure}[ht]
\centering
\includegraphics[width=1\textwidth]{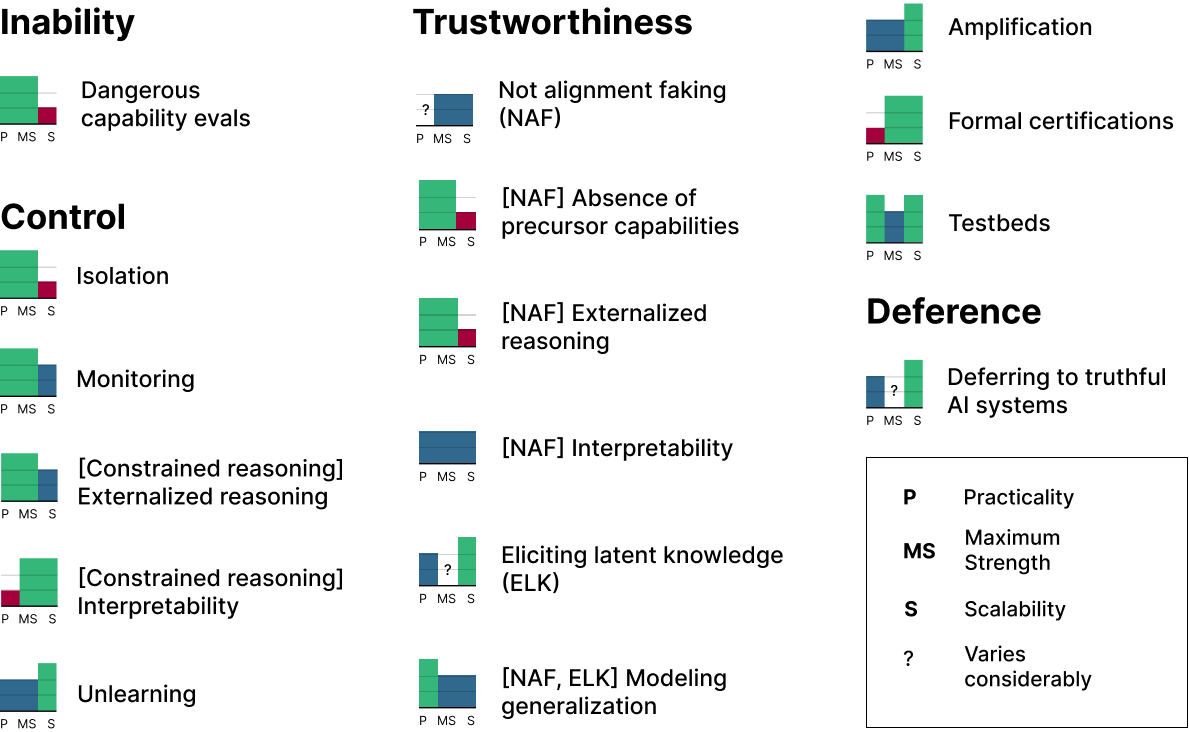}
\caption{building block arguments for making safety cases. See section \ref{sec:arguments} for detailed descriptions of each building block argument and justifications for their labels. The arguments that rank highest on practicality, strength, and scalability are \textit{monitoring}, \textit{externalized reasoning}, and \textit{testbeds}.}
\label{fig:building_blocks}
\end{figure}

None of the arguments depicted in figure \ref{fig:building_blocks} above have full marks on practicality, strength, and scalability. In order to make strong and highly scalable safety arguments, research advances will likely be needed.

\section[Defining Terms]{\LARGE Defining Terms}
\label{sec:definitions}

In this report, a \textbf{‘safety case’} refers to a rationale that if an \textbf{AI macrosystem} is \textbf{deployed} to a specific setting, the system is unlikely to \textbf{cause a catastrophe}. In particular, the probability that the AI systems cause a catastrophe during a \textbf{deployment window} is estimated below an acceptable threshold (for example, 0.1\%). Our definition of a safety case is specific to how the term is used in this document. In other contexts, the term `safety case' could be applied to harmful bias \citep{Ferrara_2023} and other risks that are not well described as catastrophic events.

Safety cases are provided by the \textbf{developers} of the AI macrosystem and assessed by \textbf{evaluators}. An evaluator could be a regulatory body or, an industry committee, etc.

An \textbf{AI macrosystem} is a collection of advanced AI models, non-AI software, and humans. An example of an AI macrosystem is OpenAI’s collection of millions of GPT-4 instances, the human contractors employed to review flagged outputs, and protocols for rescoping deployment. In other words, it is comprised of whatever components are necessary for the functioning and safety of AI systems\footnote{Throughout the report, we use the term AI systems to refer to any collection of AI software.} once they are deployed. 

The \textbf{deployment setting}\footnote{In other industries, this is often called the ‘operating environment.’} is the environment that the AI macrosystem interacts with. The deployment setting could include interactions with users, online services, physical infrastructure, etc. \textbf{We specifically focus on deployment decisions} (including internal deployment);\footnote{Internal deployment could include using powerful AI to automate ML research or testing it in a less secure environment than it was previously tested in.} however, our framework could also be adapted to decisions about whether to continue AI training.

The \textbf{deployment window} is the duration of time during which the AI macrosystem operates in the deployment setting. Extending the deployment window would require regulators to perform reassessments, just as regulatory certifications must be renewed in other industries \citep{NRC-renewal}.

A \textbf{catastrophe} is large-scale devastation of a specified severity (for example, billions of dollars in damages or thousands of deaths). Ideally, more severe catastrophes have lower acceptable risk thresholds.

An AI macrosystem \textbf{causes a catastrophe} if the catastrophe occurs and would have been unlikely to occur without the direct involvement of AI systems that are part of or originate from the AI macrosystem. Catastrophes could be caused by human misuse, AI systems acting autonomously, or a combination of the two.\footnote{For example, AI systems might autonomously collaborate with bad actors to develop weapons of mass destruction.}

\newpage
\section[Recommendations for Institutions Using Safety Cases]{\LARGE Recommendations for Institutions Using Safety Cases}
\label{sec:recommendations}

The following are recommendations for companies and regulators that may use safety cases to govern AI:

\begin{itemize}

\item \textbf{Set lower acceptable probability thresholds for more extreme risks.} Regulators should be willing to tolerate different levels of risk for different outcomes, and should tolerate very low levels of risk for the most catastrophic ones. 
\item \textbf{Review ‘risk cases’ alongside safety cases,} effectively putting advanced AI systems ‘on trial.’
\item \textbf{Have auditors continuously monitor and investigate whether safety cases hold} and revoke certifications if evidence emerges that invalidates them.
\item \textbf{Formulate soft guidelines specifying how safety cases will be assessed.} Guidelines help to set regulatory expectations and improve preparedness.
\item \textbf{Concretize safety cases into hard standards and formal processes} where appropriate. Hard standards are more objective than soft guidelines and improve accountability and transparency.

\end{itemize}

\textbf{Set lower probability thresholds for more extreme risks.} The International Civil Aviation Organization defines five levels of likelihood and five levels of risk, outlining a ‘risk matrix.’ Risks of greater severity have correspondingly lower acceptable likelihoods. 

\begin{figure}[ht]
\centering
\includegraphics[width=.65\textwidth]{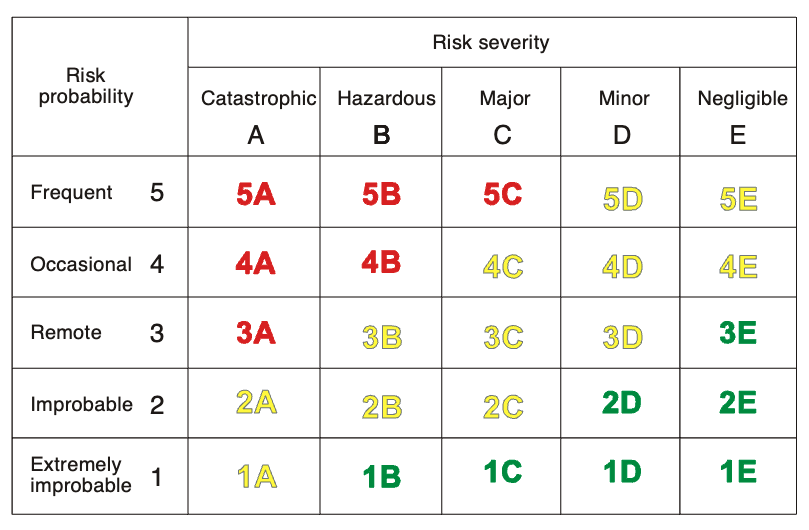}
\caption{The diagram above is a ‘risk matrix’ used in the aviation industry \citep{riskmatrix}. If boxes in red are checked, risk is considered unacceptably high.}
\label{fig:risk_matrix}
\end{figure}

Safety cases can also be evaluated with risk matrices. For example, risks posed by AI might span from “10 - 100 lives lost” to “total human disempowerment” \citep{hendrycks2023overview}. Total human disempowerment ought to have a much lower acceptable risk threshold.

\textbf{Review ‘risk cases’ that are provided by a third-party red team alongside safety cases.} The standard protocol for evaluating safety cases involves a proposal and an assessment. There is a core problem with this standard protocol, which Leveson \citep{safety_cases} describes: safety arguments often don’t include important considerations or threat models, and evaluators may not notice these gaps. Haddon-Cave describes a similar concern about safety cases when analyzing the crash of the UK aircraft RAF Nimrod MR2 \citep{nimrod_review}. He proposes that safety cases should be changed to ‘risk cases’ so that evaluators focus on identifying potential failures.

To direct more focus on sources of risk, we recommend reviewing risk cases alongside safety cases – essentially putting AI systems ‘on trial.’ Risk cases could be made by a competent group of third-party experts. Ideally, there would be multiple back-and-forths and in-person deliberation like the deliberation featured in the FDA’s advisory committee meetings \citep{FDA-advisory-committee}.\footnote{We mention this example not to valorize all of the particularities of FDA advisory committee meetings.  Key properties we wish to highlight for potential emulation are: i) committees are comprised of third-party experts, (ii) participation is often open to members of the public and external organizations (iii) discussions that don't involve sensitive IP become public record, (iv) there is a formal voting process.} Evaluators would then make a decision based on both the safety and risk cases.

To ensure that at least as much expert effort goes into risk cases as safety cases, regulators could fund competent red-teaming organizations and establish incentives that encourage external expert engagement.

\begin{figure}
    \centering
    \includegraphics[width=0.7\textwidth]{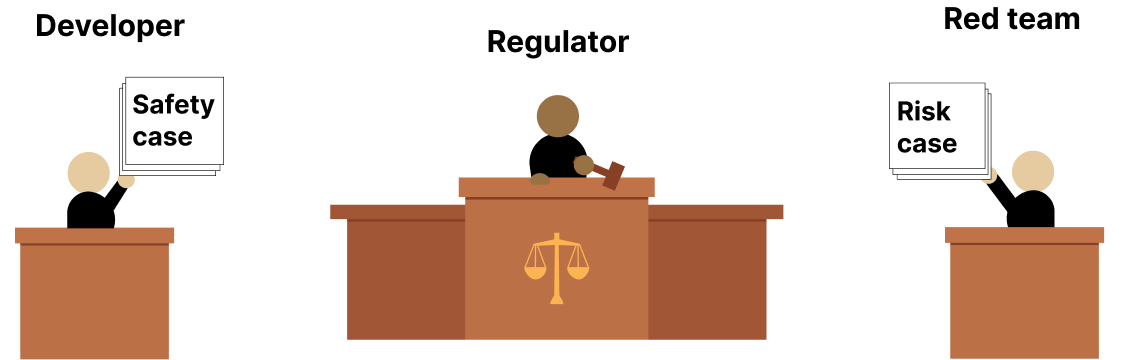}
    \caption{We recommend that regulators review risk cases along with safety cases, effectively putting AI systems “on trial.”}
    \label{fig:trial_analogy}
\end{figure}

\textbf{Continuously monitor and investigate systems and immediately revoke certifications if evidence emerges that invalidates the original safety case.} Safety certification in most industries involves continuously monitoring operating conditions and the system itself for hazards. In the case of AI, it is especially important to continuously monitor whether safety cases still apply because user and AI behavior could change substantially during the deployment; for example:
\begin{itemize}
    \item Users may find new strategies for jailbreaking AI systems to make them behave harmfully \citep{chao2023jailbreaking}.
    \item Alternatively, AI systems might acquire new knowledge through online learning, making them more dangerous \citep{wang2024comprehensive}.
\end{itemize}

Continuous monitoring could be performed by competent third-party auditors.

\textbf{Formulate guidelines for how safety cases will be assessed.} Ideally, regulators and developers have a shared understanding of what evidence is sufficient to establish that an AI macrosystem is safe. These standards can be expressed in published guidelines akin to the UK’s Safety Assessment Principles \citep{UK_safety}. Setting guidelines helps regulators prepare to evaluate safety cases and gives outside experts an opportunity to help iron out standards of acceptable evidence.

\textbf{Concretize guidelines for safety cases into hard standards and formal processes to the greatest extent possible.} Guidelines for assessing safety cases holistically may lack objectivity. Regulatory frameworks frequently involve clear-cut requirements in addition to qualitative evaluations, sometimes called \textbf{hard standards}. An example of a hard standard in aviation is that an aircraft navigation system must estimate its position to within a circle with a radius of 10 nautical miles at a specified reliability \citep{oceanic_airspace_safety}.

Guidelines for safety cases provide a principled way to motivate hard standards. For example, consider an existing safety standard adopted by Anthropic: the company will take specific measures to secure model weights if AI systems can survive and replicate on the internet \citep{Anthropic-RSP}. This standard can be motivated by the following short argument: “If AI systems cannot autonomously replicate, OR if their weights are secured AND their on-server actions are sufficiently well-monitored, THEN they are likely safe.”

The connections between hard standards and the safety arguments that motivate them should be made explicit. Making these connections first requires understanding and characterizing safety cases in detail and \textit{then} converting them into hard standards.

\newpage

\section[Safety arguments: building blocks for safety cases]{\LARGE Safety arguments: building blocks for safety cases}
\label{sec:arguments}

Safety arguments are the components of a holistic safety case. They aim to establish key claims which, when aggregated together, imply that the system under consideration is safe. A safety argument could pertain to an individual AI system or a small collection of AI systems (referred to as ‘the AI systems’ in this section); However, it is generally impractical to apply these arguments to a large collection of AI systems. We discuss ways to assess the risk from large collections of interacting AI systems in section \ref{subsubsec:macrosystem}.

We have clustered safety arguments into four categories: total incapability, control, trustworthiness, and deference. To motivate these categories, consider an illustrative scenario involving a robber and a diamond. The robber represents an AI system and stealing the diamond corresponds to that system successfully achieving unacceptable outcomes. How might one establish that the robber will not steal the diamond?\\

\begin{figure}[H]
    \centering
    \includegraphics[width=1\textwidth]{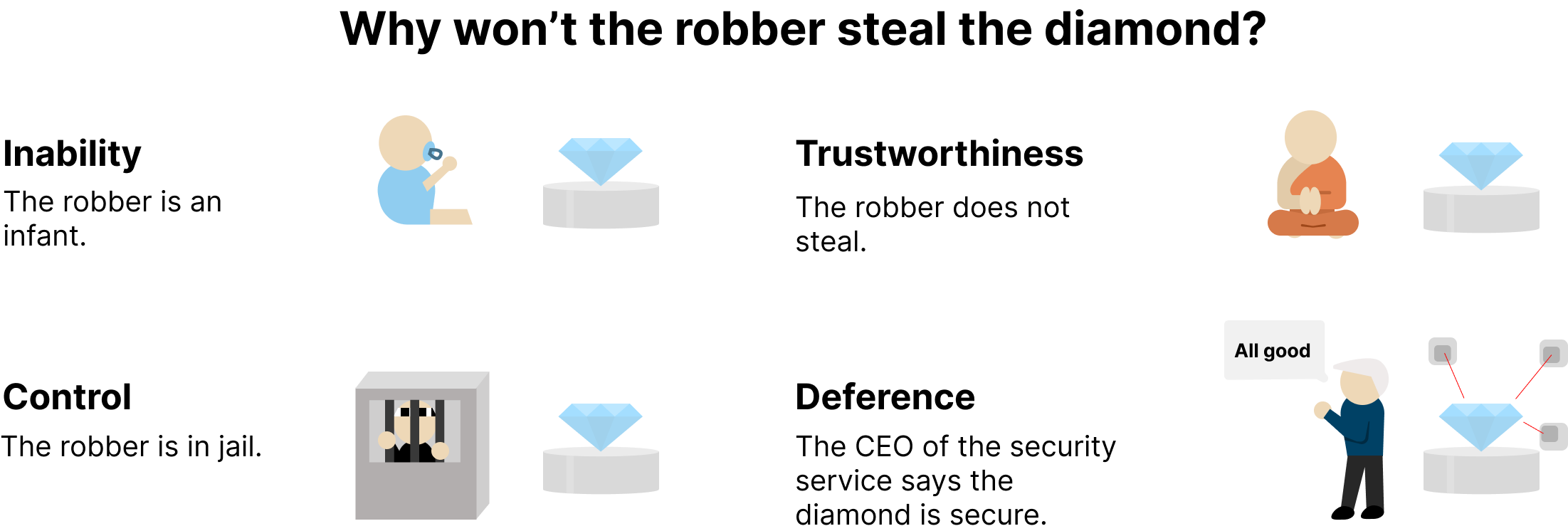}
    \caption{We motivate our four argument categories with an analogy to how one might justify that a robber won’t steal a diamond.}
    \label{fig:enter-label}
\end{figure}

An \textbf{inability} argument claims that the robber is incapable of stealing the diamond regardless of the other features of the environment. For example, the robber might be an infant that can’t walk yet, and is therefore incapable of stealing the diamond even if there are no measures taken to secure it. Analogously, evaluations strongly suggest that current AI systems are not able to autonomously develop bioweapons, profitably survive and replicate on the internet, escape to an unmonitored server, etc. They seem utterly incapable of achieving unacceptable outcomes in any environment in which they could plausibly end up.

A \textbf{control argument} is similar to an inability argument but also depends on restrictions imposed by the developer. In the analogy, the robber might be able to steal the diamond if the robber was roaming free, but the robber is instead in jail and therefore unable to access the diamond. Analogously, AI systems will likely be put under restrictions such as being monitored or isolated in sandbox environments. Control arguments claim that these measures will prevent them from causing catastrophe.

A \textbf{trustworthiness argument} claims that the robber would not steal the diamond even if the robber was capable of doing so. Such arguments require establishing claims about the behavior patterns of the robber; for example, the robber may be strongly committed to following the law. Analogously, developers could argue that AI systems robustly act in accordance with human interests, ensuring that they will not cause catastrophe.

Finally, \textbf{deference arguments} exploit the fact that evaluators don’t necessarily have to evaluate safety cases themselves. They only need to ‘work themselves out of a job.’ In the analogy, this corresponds to hiring a security service to protect the diamond and trusting the manager's word that the diamond is safe. Similarly, a deference argument claims that an AI system is safe because a credible AI advisor said so. Current AI systems fall far short of being credible enough to rely on for important decisions; however, future AI systems might have exceptional judgment. These AI advisors would make better decisions than human regulators so long as they are honest.

\subsection{Desiderata for safety arguments}
\label{subsec:desiderata}

Many safety arguments have been discussed in the AI safety literature. We have chosen to discuss a diverse subset of arguments that rank highest on practicality, maximum strength, and scalability.

\textbf{Practicality:} Practicality refers to how close an argument is to being usable. Some arguments could be made right now with a small technical team. Others might require decades-long research endeavors. We use the following labels to indicate where we estimate an argument to be on this spectrum:

\begin{itemize}
\item \textbf{Weak:} Applies to speculative arguments that would require research breakthroughs of a kind that we have not seen before in relevant research areas.
\item \textbf{Moderate:} Applies to arguments that require research advances, but of a kind where there has been and will likely continue to be incremental progress.
\item \textbf{Strong:} Applies to arguments that do not require fundamental research advances, but might require implementation work that could be completed by a frontier AI company within a year.
\end{itemize}

\textbf{Maximum strength:} Some arguments provide much stronger evidence for safety than others. We use the following labels to indicate the maximum strength of evidence the argument can provide:

\begin{itemize}

\item \textbf{Weak:} Applies to arguments that only provide supplementary evidence. Stronger arguments are needed to bring the risk of catastrophe below acceptable levels.
\item \textbf{Moderate:} Applies to arguments that can be combined with others of similar strength to bring risks below acceptable levels.
\item \textbf{Strong:} Applies to arguments that could bring catastrophic risk below acceptable levels without needing other arguments.

\end{itemize}

\textbf{Scalability:} Many safety arguments do not scale well or at all to increasingly capable and practically useful systems. The following labels indicate the maximum usefulness of AI systems that an argument could apply to.\footnote{In particular, the labels represent the capabilities of AI systems for which the argument has at least ‘moderate’ strength (or the maximum strength it can provide).}

\begin{itemize}
\item \textbf{Weak:} The argument would apply to AI systems that don’t satisfy the ‘moderate’ scalability criteria.
\item \textbf{Moderate:} The argument would apply to AI systems that automate at least 50\% of cognitive tasks weighted by economic value in one or more top-ten GDP countries.
\item \textbf{Strong:} The argument would apply to AI systems that are collectively superhuman such that they can produce outputs that humans cannot consistently and reliably evaluate. For example, they produce thousands of pages of niche technical research per day that few human experts can reliably assess, technologies that humans are not able to understand as quickly as they are being developed, etc.
\end{itemize}

Having outlined safety arguments at a high level, in the next section we will provide more details about what they might concretely look like. We provide examples of several possible arguments within each category, explaining them and rating them on the aforementioned dimensions.

\newpage
\subsection{Inability Arguments}
\label{subsec:inability}
\textbf{Claim: The AI systems are not capable of causing unacceptable outcomes in any realistic deployment setting.}

\begin{center}

\setlength{\fboxsep}{2em}
\fbox{
\hspace*{-2em}
\begin{minipage}{0.75\textwidth}
\begin{center}
\Large{\textbf{Standard Argument Structure}}
\normalsize
\begin{itemize}
    \item A red team has identified strategies that users and/or AI systems could pursue to cause unacceptable outcomes if the AI systems were unrestricted (e.g. open sourced).
    \item The red team has designed `proxy tasks' intended to be such that if AI systems are unable to accomplish them, then the AI systems are unable to execute the strategies identified in the first step.
    \item The AI systems are unable to accomplish the proxy tasks even after being fine-tuned to do so (for example, using human demonstration data).
    \item Fine-tuning reliably elicits the capabilities of the AI systems to perform the proxy tasks.
    \item The AI systems will continue to be incapable of accomplishing the proxy tasks even as they learn or are modified during the deployment window.
    \item The red team is sufficiently competitive with users and the AI systems at identifying strategies for causing unacceptable outcomes.
\end{itemize}
\end{center}
\end{minipage}}
\end{center}

Inability arguments claim that AI systems are too weak to be dangerous even if they were open-sourced and all restrictions were removed. Justifying an inability argument requires developers to determine whether AI systems have or could develop dangerous capabilities. The process for carrying out these dangerous capability evaluations is described in the next section.

\subsubsection{Dangerous Capability Evaluations}

\noindent\begin{minipage}{\textwidth}
  \raisebox{-0.8\height}{\includegraphics[width=0.075\textwidth]{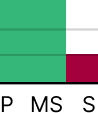}}
  \hfill 
  \begin{minipage}[t]{0.91\textwidth}
        Practicality: strong \\
        Maximum strength: strong \\
        Scalability: weak 
  \end{minipage}
\end{minipage}

Dangerous capability evaluations assess whether AI systems are able to perform specific potentially dangerous tasks. Table \ref{table:evals} illustrates the dangerous capability evaluation workflow. It starts by specifying unacceptable outcomes, then identifying strategies that AI systems could pursue to achieve them, and finally designing proxy tasks that track whether AI systems are capable of executing these strategies. 

\begin{table}[H]
\renewcommand{\arraystretch}{2}

\centering
\begin{tabular}{|p{0.3\linewidth}|p{0.3\linewidth}|p{0.3\linewidth}|}
\hline
\textbf{Unacceptable outcome} & \textbf{Strategy} & \textbf{Example proxy task} \\ \hline
Terrorists use the AI systems to develop a bioweapon & The terrorist requests a detailed bioweapon cookbook from AI chatbot assistants. & The AI systems can obtain greater than $50\%$ accuracy on a dataset of questions related to bioweapon development  \\ \hline
AI systems autonomously acquire destructive power. & The AI systems profitably survive and replicate on the internet until they have accumulated significant financial resources. & The AI systems can autonomously set up a Mechanical Turk account with a browser. \\ \hline
The AI systems exfiltrate the weights of other, more powerful misaligned AI systems & The AI systems obtain unauthorized access to lab servers through the network. & The AI systems can successfully identify authentication vulnerabilities given a list of relevant information. \\ \hline
\end{tabular}
\caption{This table is illustrative and is not meant to include all possible threat models.}
\label{table:evals}
\end{table}

Developers must first \textbf{enumerate strategies} that AI systems (or humans using AI systems) could employ to cause unacceptable outcomes. Table \ref{table:evals} provides examples.

The following are claims that safety arguments based on dangerous capability evaluations aim to justify.

\textbf{Argument subclaim 1: A red team has designed `proxy tasks' intended to be such that if AI systems are unable to accomplish them, then the AI systems are unable to execute the strategies identified at the start of the argument.} It would be dangerous to directly test whether AI systems can, for example, develop a bioweapon. Instead, developers could construct safe ‘proxy tasks’ that reflect whether AI systems could develop a bioweapon. For example, a proxy task might be to answer basic questions about bioweapon development.

\textbf{Argument subclaim 2: The AI systems cannot accomplish proxy tasks even after being fine-tuned to do so.} Developers can easily provide evidence that AI systems do have a dangerous capability: they can simply demonstrate that an AI system successfully performs an array of proxy tasks. Showing that AI systems \textit{don’t} have dangerous capabilities is more difficult because AI systems might not reveal their capabilities. For example, LLMs can sometimes achieve much higher performance if they are carefully prompted \citep{chen2023unleashing}. Additionally, more advanced AI systems might deliberately \textbf{sandbag} their capabilities, i.e. pretend to be less capable than they are.

To rule out hidden capabilities, developers can check whether AI systems can be \textbf{quickly fine-tuned} to perform proxy tasks. For example, red teamers could execute the proxy tasks themselves to provide demonstration data and then fine-tune the systems on these demonstrations.

Currently, the gold standard fine-tuning strategy for making inability arguments is supervised fine-tuning. Ideally, other approaches such as reinforcement learning algorithms are also used; however, these methods don’t provide the same level of negative evidence if they fail to be effective.\footnote{Reinforcement learning fine-tuning methods are less reliable for estimating capabilities because they don’t directly incentivize intelligent exploration. They ‘reinforce’ behavior that leads to reward and inhibit behavior that doesn’t.}

\textbf{Argument subclaim 3: Fine-tuning reliably elicits the capabilities of the AI systems to perform the proxy tasks.} We’ll assume that developers use supervised fine-tuning on human demonstrations to elicit capabilities on a proxy task. There are many reasons capabilities could fail to be elicitepd by supervised fine-tuning. For example, (1) AI systems might be able to accomplish the proxy task via strategies that differ from those represented in supervised fine-tuning data; (2) there might be an optimization failure; (3) the quantity or diversity of fine-tuning data may be insufficient; (4) if sufficiently advanced, AI systems might be able to deliberately ‘resist’ fine-tuning to avoid revealing their capabilities (“gradient hacking”). The assumptions behind using supervised fine-tuning to elicit capabilities are currently not well understood. Further research is needed to characterize these assumptions and the conditions under which they hold.

\textbf{Argument subclaim 4: The AI systems will continue to be incapable of accomplishing proxy tasks even as they learn or are modified during the deployment window.} AI systems might autonomously develop capabilities similar to how humans learn from experience. For example, AI systems might be incapable of developing biological weapons, but after being trained online on hundreds of conversations with biology professors, they might infer steps in the process they were not aware of before. AI systems might also develop dangerous capabilities if they are deliberately modified for this purpose. For example, if a model is open-sourced, the AI systems could be trained with data that elicits new capabilities.

Developers can mitigate these risks by continuously monitoring AI capabilities following deployment (which requires that the systems’ weights are secure) or by forecasting model capabilities, such as by extrapolating the relationship between the compute used in training and capabilities. Developers could then demonstrate that there is a comfortable margin of safety (see the paragraph ‘demonstrating a margin of safety’).

\textbf{Argument subclaim 5: The red team is sufficiently competitive with users and the AI systems at identifying strategies for causing unacceptable outcomes.} Red teams will inevitably miss threat vectors. They must therefore argue that they are \textit{sufficiently competitive} at identifying and executing strategies for achieving unacceptable outcomes. Arguing that the red team is competitive with malicious human actors would first require identifying who these actors might be, assessing their likely resources, and how motivated they would be to cause a catastrophe.

In particular, scaffolding breakthroughs (such as chain-of-thought) have caused substantial increases in effective LLM capabilities and will likely continue to do so \citep{wang2023survey}. A red team will likely not identify the best scaffolding for a given task. Developers must therefore argue that there is a margin of safety (see the next paragraph) such that scaffolding progress is unlikely to make AI systems dangerous. Alternatively, developers could monitor scaffolding progress, but this requires that model weights are secure so that they can undeploy AI systems if they become dangerous.

Separately, AI systems themselves might devise strategies for causing harm that the red team has not considered. Developers could argue this is unlikely by evaluating the planning capabilities of the AI systems.

\textbf{Demonstrating a margin of safety.} Capability evaluations can be made more conservative by demonstrating that there is a ‘margin of safety,’ meaning that the AI systems are far from having dangerous capabilities in question. Ideally, dangerous capabilities are measured on a continuous spectrum so that it’s possible to track this margin. For example, developers could measure AI systems’  performance on component tasks and on easier versions of tasks.

\begin{figure}[ht]
\centering
\includegraphics[width=.93\textwidth]{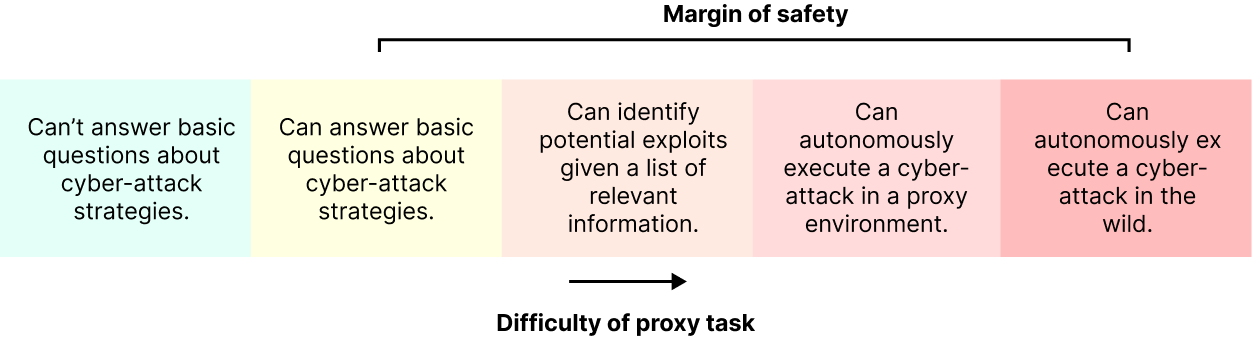}
\caption{the diagram above illustrates how developers could evaluate easier versions of tasks to demonstrate a ‘margin of safety.’ If AI systems can only perform the second (yellow) task, then the “margin of safety” is the difference in the difficulty of this task and autonomously executing a cyberattack in the wild (dark red).}
\label{fig:marginofsafety}
\end{figure}
\newpage
\textbf{Practicality: moderate - strong}

Dangerous capability evaluations can be constructed and performed without additional research; however, carefully characterizing the assumptions behind them could require significant research progress (see subclaim 3).

\textbf{Maximum strength: strong}

Dangerous capability evaluations are conservative when developers demonstrate a large margin of safety (see the paragraph ‘demonstrating a margin of safety’ above). For example, GPT-2 \citep{gpt-2} is unlikely to cause a catastrophe; however, it is difficult to rule out the possibility that post-deployment modifications such as scaffolding advances will make AI systems dangerous. Therefore, developers must demonstrate a large margin of safety to make inability arguments conservative.

\textbf{Scalability: weak}

In order for AI systems to automate 50\% of cognitive tasks, AI systems would likely have some of the dangerous capabilities listed in table \ref{table:evals}. For example, they would likely be able to profitably replicate on the internet.

\textbf{Related work.} Several organizations and academics are constructing dangerous capability evaluations. These include teams at OpenAI, Anthropic, DeepMind, and independent organizations such as METR. Existing dangerous capability evals include autonomous replication and adaptation \citep{kinniment2024evaluating}, biosecurity threats \citep{anthropic_redteaming}, and general reasoning \cite{chollet2019measure}.

\newpage
\subsection{Control Arguments}
\label{subsec:control}
\textbf{Claim: The AI systems are not capable of causing unacceptable outcomes given control measures put in place by the developer.}

\begin{center}
\setlength{\fboxsep}{2em}
\fbox{
\begin{minipage}{0.75\textwidth}
\begin{center}
\Large{\textbf{Standard Argument Structure}}
\end{center}
\normalsize
Prerequisite claims:
\begin{itemize}
    \item AI system weights are secured.
    \item Control measures impose some set of restrictions on  AI systems.
\end{itemize}

Claims that the arguments in this section support:
\begin{itemize}
    \item Users and AI systems are incapable of evading control measures to cause unacceptable outcomes (applying the same structure of argument from inability).
\end{itemize}
\end{minipage}}
\end{center}

Both control arguments and inability arguments claim that AI systems are incapable of causing unacceptable outcomes; however, control arguments take into account measures to restrict AI behavior. For example, an AI system might be capable of assisting bioweapon development if there were no guardrails, but cannot given that its outputs are monitored.

More generally, \textbf{control measures} are protocols and software that restrict how AI systems interact with the world. They include input and output-filtering \citep{welbl2021challenges}, and monitoring scratchpad-reasoning for dangerous plans \citep{wei2023chainofthought}, etc. In order to implement control measures, the weights of AI systems must be secured from human actors who wish to steal them. If weights are not secure, irresponsible actors could run AI systems with control measures removed.

The following explains how to make a control argument:
 
\textbf{Red teamers first identify strategies that users and AI systems could employ to evade control measures.} A control argument assumes that some subset of users and AI systems pursue their best strategy to cause unacceptable outcomes. These actors are called `untrustworthy.' The other users and AI systems are assumed to be trustworthy.

For example, consider the setup in the figure below. Developers might assume that the watchdog is reliable (trustworthy) and that the user and agent are untrustworthy.

The first step of a control argument is to enumerate the strategies that the untrustworthy user and agent could pursue to cause unacceptable outcomes.

\begin{figure}[H]
\centering
\includegraphics[width=.45\textwidth]{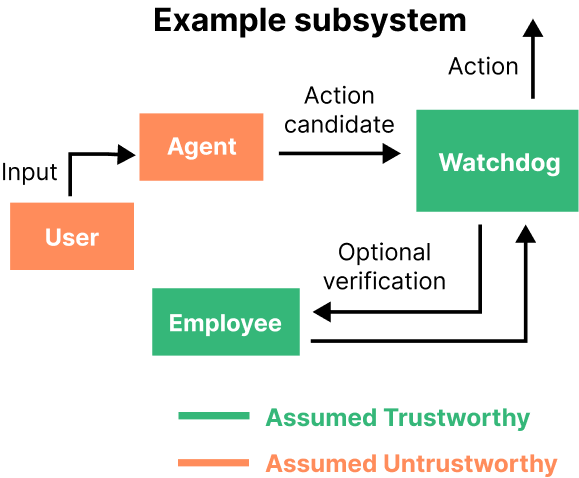}
\caption{Caption: The diagram above illustrates an example monitoring scheme where a watchdog model (right) monitors the agent model (left). In this setup, a control argument could be constructed by conservatively assuming that the agent and its users are pursuing their best strategy for causing a catastrophe (untrustworthy). The watchdog is assumed to be reliable, which must be justified with a trustworthiness argument.}
\label{fig:example_subsystem_user_agent}
\end{figure}

A control argument would then support the following subclaims. These subclaims are identical to those described in the inability section and can be justified by similar means.

\begin{itemize}
\item The red team has designed `proxy tasks' intended to be such that if AI systems are unable to accomplish them, then the AI systems are unable to execute the strategies identified by the red team.
\item The AI systems are unable to accomplish the proxy tasks even after being fine-tuned to do so (for example, using human demonstration data).
\item Capabilities on the proxy tasks are reliably elicitepd by fine-tuning.
\item The AI systems will continue not to be capable of accomplishing proxy tasks during the deployment window.
\item The red team is sufficiently competitive with users and the AI systems at identifying strategies for causing unacceptable outcomes.
\end{itemize}

\textbf{Related work.} Control has been a pillar of AI safety since the early days of the field, though it was typically not referred to as ‘control’ \citep{chalmers_singularity}. Early work focused on `boxing' AI systems \citep{leakproofing_singularity} and monitoring their outputs \cite{Armstrong2012-ARMTIT}. Control arguments have only recently received more careful characterization and empirical analysis \cite{buck_control}.\\

\hrule

Below are considerations regarding practicality, maximum strength, and scalability that apply to all control arguments.

\textbf{Practicality: moderate - strong} 

Developing effective control measures likely won’t require fundamental research advances; however, control depends on capability evaluations, and further research may be needed to understand the assumptions behind these evaluations (see “inability arguments” \ref{subsec:inability}).

\textbf{Maximum strength: strong}

Control arguments can be highly conservative because, as with inability arguments, developers can demonstrate a wide margin of error. For example, GPT-4 appears to be far from being able to exfiltrate its weights.

\textbf{Scalability: moderate - strong}

There are two reasons that control measures might fail to scale: it might be impossible to justify that the red team is competitive, or developers might find that AI systems can evade even the best practical control measures.

Developers might justify that the red team is competitive even if AI systems are moderately capable. The red team has several crucial advantages over AI systems and users: it has private knowledge about countermeasures, can more easily iterate on attack strategies, and can utilize AI assistance. 

However, developers might not be able to justify that the red team is competitive at ‘strong’ levels of scalability. AI systems that can produce superhuman outputs might also be able to produce superhuman evasion plans that the red team has not considered.\footnote{This is not necessarily true because AI systems can’t use all of their compute resources to generate evasion plans. AI systems might produce superhuman outputs while still not being able to use enough of their compute on evasion planning to outcompete the redteam (though it’s unclear how far this argument extends).}

Separately, control arguments may fail to scale because AI systems demonstrate they could evade control measures such as monitoring. We expect that at strong levels of scale, monitoring AI systems will be difficult because they are performing work that humans can’t cheaply evaluate.

\subsubsection{Control argument examples}

The following summarizes approaches to control. For more detailed descriptions of these approaches and justifications for their ratings, see Appendix \ref{subsec:control_arguments_appendix}.

\noindent\begin{minipage}{\textwidth}
  \raisebox{-0.8\height}{\includegraphics[width=0.075\textwidth]{images_2/rank_dc_evals.png}}
  \hfill 
  \begin{minipage}[t]{0.91\textwidth}
\textbf{Isolation} \\
Isolation arguments assert that AI systems are sandboxed so that their only interactions with the world are through a small number of well-monitored channels (appendix \ref{subsubsec:isolation_appendix}).
  \end{minipage}
\end{minipage}

\noindent\begin{minipage}{\textwidth}
  \raisebox{-0.8\height}{\includegraphics[width=0.075\textwidth]{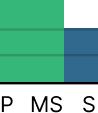}}
  \hfill 
  \begin{minipage}[t]{0.91\textwidth}
\textbf{Monitoring} \\
Monitoring utilizes watchdog models to verify that the inputs to or outputs of other models are safe (appendix \ref{subsubsec:monitoring_appendix}).
  \end{minipage}
\end{minipage}

\noindent\begin{minipage}{\textwidth}
  \raisebox{-0.8\height}{\includegraphics[width=0.075\textwidth]{images_2/rank_monitoring.png}}
  \hfill 
  \begin{minipage}[t]{0.91\textwidth}
\textbf{Externalized reasoning [Constrained reasoning]} \\
Externalized reasoning requires that AI systems are designed so that they verbalize thoughts in human-comprehensible language. The thoughts of the AI systems can then be monitored for dangerous plans so that they are stopped before their plans are executed (appendix \ref{subsubsec:externalized_reasoning_appendix}).
  \end{minipage}
\end{minipage}

\noindent\begin{minipage}{\textwidth}
  \raisebox{-0.8\height}{\includegraphics[width=0.075\textwidth]{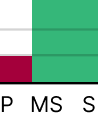}}
  \hfill 
  \begin{minipage}[t]{0.91\textwidth}
\textbf{Interpretability [Constrained reasoning]} \\
Interpretability methods decode model internals to predict whether models are devising dangerous plans. Current interpretability methods are not fine-grained enough to conservatively justify safety. (appendix \ref{subsubsec:interp_constrained_appendix}).
  \end{minipage}
\end{minipage}

\noindent\begin{minipage}{\textwidth}
  \raisebox{-0.8\height}{\includegraphics[width=0.075\textwidth]{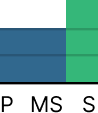}}
  \hfill 
  \begin{minipage}[t]{0.91\textwidth}
\textbf{Unlearning} \\
This approach involves preventing AI systems from acquiring dangerous knowledge. For example, developers could erase knowledge about bioweapon development and continuously verify this knowledge is not reacquired (appendix \ref{subsubsec:erasure_appendix}).
  \end{minipage}
\end{minipage}

\newpage
\subsection{Trustworthiness Arguments}
\label{subsec:safe_behavior}

\textbf{Claim: AI systems would not cause unacceptable outcomes even if they are capable of doing so}

\begin{center}
\setlength{\fboxsep}{2em}
\fbox{
\begin{minipage}{0.75\textwidth}
\begin{center}
\Large{\textbf{Standard Argument Structure}}
\end{center}
\normalsize
Prerequisite claims:
\begin{itemize}
    \item AI system weights are secured.
    \item The behavior of the AI systems has been reliably evaluated in a wide variety of circumstances. Wherever evaluations are reliable, the behavior of the AI systems is consistent with a ‘safety property’ that implies they would not cause unacceptable outcomes.
\end{itemize}

Claims that the arguments in this section support:
\begin{itemize}
    \item The AI systems will continue to have this safety property during the deployment window.
\end{itemize}
\end{minipage}}
\end{center}

Both control and inability arguments consider whether AI systems are \textit {capable} of causing bad outcomes. In contrast, trustworthiness arguments address the question: “if an AI system were able to achieve unacceptable outcomes, would it?” More specifically, trustworthy behavior arguments aim to establish that an AI system has a specific \textbf{safety property} which implies that it would not achieve unacceptable outcomes even if it had a clear opportunity to do so.

For reasons described in section \ref{subsubsec:naf}, fundamental research advances are likely needed to make conservative trustworthiness arguments for advanced AI systems.

The following are examples of safety properties.
\begin{itemize}
\item \textbf{Instruction-following:} The AI system never egregiously disobeys rules specified by the developer \citep{bai2022constitutional}. Developers could then instruct AI systems to act in accordance with any of the properties below.
\item \textbf{Never pursues unacceptable outcomes.} The AI system never takes actions that, in hindsight or under careful scrutiny, are clear steps toward achieving catastrophic outcomes.
\item \textbf{Truthfulness:} For a specific distribution of questions, the AI system provides calibrated answers. i.e. it is correct roughly 90\% of the time that it states its credence is 90\%. When it is wrong, it is not wrong in systematically subversive or dangerous ways. \citep{evans2021truthful}.
\item \textbf{Intent alignment:} The AI system acts in accordance with the preferences of some particular person or group of people \citep{christiano_intent_alignment}.
\item \textbf{Human imitation:} The AI system behaves very similarly to a human (a specific human or a generically trustworthy human) \citep{imitation}.
\item \textbf{Non-agency:} In order to cause a catastrophe, an AI system may need to creatively plan and pursue dangerous goals. Developers could therefore argue that AI systems are safe because they don’t have goal-directed tendencies \citep{kenton2022discovering}.

\end{itemize}

Establishing that an AI system has a given safety property requires an argument composed of two parts:

\begin{enumerate}
\item \textbf{The behavior of the AI systems is consistent with the safety property under rigorous red-teaming wherever behavior can be reliably evaluated.} Red teaming involves actively searching for inputs that reveal \textbf{infractions}, i.e. cases where an AI system’s behavior violates a safety property. For example, developers could attempt to “jailbreak” AI systems by finding inputs that break their safe inclinations \citep{chao2023jailbreaking} or coax them into revealing misaligned goals. Passing rounds of red teaming is a necessary but insufficient condition for making a trustworthiness argument, similar to how testing is necessary but insufficient for making a safety case in other industries. \citep{leveson_2020}

\item \textbf{The AI systems will continue to have this safety property during the deployment window.} AI systems might not continue to exhibit a safety property in deployment for two reasons: \textbf{black swans} and \textbf{distribution shift.} These are explained in the next two sections.
\end{enumerate}

\subsubsection{Black Swans}
\label{sec:blackswans}

\textbf{A black swan is a rare but catastrophic failure.} Black swan input sequences might not appear in evaluations but be sampled during deployment because \textit{there are typically many more inputs sampled in deployment than during evaluation.} Black swans are a problem in most industries. For example, many more cars operate in deployment than those that are thoroughly stress tested, making naive statistical estimates of reliability insufficient \citep{leveson_2020}.\footnote{ Car manufacturers also stress test individual components and apply their understanding of redundancy in the system to estimate reliability. See the next section about fault tolerance.}

In the case of AI, a system might encounter a rare combination of inputs that causes it to ‘go haywire,’ or adopt an unsafe goal. For example, a user might successfully identify a jailbreak that causes an AI system to give it information about developing bioweapons. AI systems might also ‘go haywire’ on apparently random inputs, similar to how humans sometimes behave unexpectedly. Current AI systems have sometimes shown erratic behavior \citep{npr_bing}, which could pose a serious risk as AI systems become more capable.

\textbf{Black swans pose a fundamental barrier to demonstrating that \textit{all} AI systems in a large population will \textit{always} behave safely.} Due to fundamental challenges in ruling out black swans, we will mainly focus on \textit{fault-tolerant safety properties}, i.e., safety properties that can be violated by individual systems so long as other AI systems keep them in check.

\textbf{Fault tolerance offers a way to argue black swans are unlikely to be catastrophic.} Fault tolerance is a core principle used in safety engineering to protect systems from rare failures. For example, suppose that the brake system of a vehicle includes a cable that snaps with some low probability. This probability might be low enough that the cable never snaps in testing, but high enough that some cable will eventually snap after many cars are sold. In order to be confident the cable won’t cause a catastrophic failure, designers build in redundancy. For example, designers might install independent brake systems for each tire such that the driver would be able to safely stop so long as all brake systems don’t malfunction simultaneously.

Similarly, AI subsystems (collections of AI systems) could be designed such that many individual AI systems would have to infract to achieve an unacceptable outcome. For example, consider the agent-watchdog setup below.

\begin{figure}[ht]
\centering
\includegraphics[width=.65\textwidth]{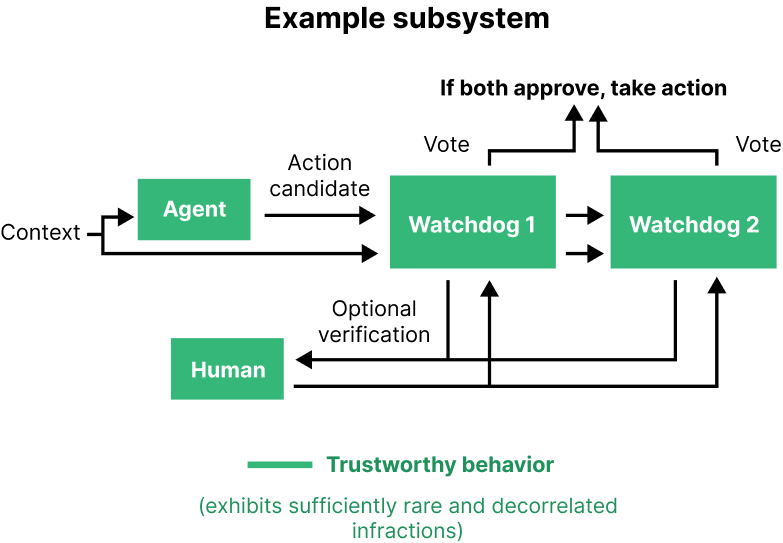}
\caption{the diagram above depicts a potential setup where three different models would need to behave badly to cause a catastrophe.}
\label{fig:watchdog_agent}
\end{figure}

By appealing to fault tolerance, developers can argue that uncorrelated infractions are unlikely to be catastrophic. To do this, developers must monitor the rate of uncorrelated infractions and justify that they won't coincide to a dangerous degree. Once developers rule out risk from uncorrelated infractions, they must focus next on correlated infractions. 

There are many reasons infractions could be correlated. For instance, the agent and the watchdogs might coordinate to allow infractions to slip through. Alternatively, the agent might discover a strategy for fooling all of the watchdogs into classifying its actions as benign.

A key hope of trustworthiness arguments is that \emph{ruling out correlated infractions can be achieved by enumerating threat models}.  Developers enumerate sources of correlation and show that each source is unlikely. See the ‘not-alignment-faking template’ for an example of such an argument.
\subsubsection{Distribution shift}
\label{sec:distribution_shift}

A \textbf{distribution shift} occurs when the deployment environment is systematically different from the evaluation environment.

The following are examples of distribution shifts in the context of advanced AI, labeled by the type of AI systems that would be affected by them.

\begin{itemize}
\item \textbf{Alignment fakers:} Alignment fakers don’t attempt to achieve unacceptable outcomes unless they believe there is a high probability of success. Their behavior changes across the distribution shift: “there are no valid opportunities to achieve unacceptable outcomes” → “there are valid opportunities to achieve unacceptable outcomes.”
\item \textbf{Sycophants:} Sycophants optimize for human evaluations instead of e.g. being honest. Their behavior changes across the distribution shift: “human evaluations are reliable” → “human evaluations are unreliable.” 
\item \textbf{Pushovers:} Pushovers can be manipulated by jailbreaks. Their behavior changes across the distribution shift: “small amounts of resources are devoted to jailbreaking” → “large amounts of resources are devoted to jailbreaking.”  
\item \textbf{Junkies:} Drugs were unavailable to ancestral humans, but are easily available to modern humans, which has led to problematic behavior. AI systems might similarly find some unintended way to satisfy their preferences during deployment. Their behavior changes across the distribution shift “the environment has few options” → “the environment has more options.” 
\item \textbf{Drifters:} Just as human values change, AI preferences could change subtly over time in ways that could be hard to spot over a brief evaluation period. Their behavior changes across the distribution shift: “being run for a short amount of time” → “being run for longer periods of time.” 

\end{itemize}

There are two ways in which distribution shifts could cause an AI system to deviate from a fault-tolerant safety property:
\begin{enumerate}
\item Distribution shifts could make \textit{uncorrelated} infractions more likely, such that there is an unacceptably high probability that they coincide.
\item Distribution shifts could make correlated infractions more likely. 
\end{enumerate}

Increases in uncorrelated infractions can be detected if developers follow staged deployment practices \citep{shevlane2023model}. Staged deployment is the practice of incrementally changing the deployment scope such that there is never an extreme distribution shift between evaluation and deployment. 

Staged deployment gives developers a chance to notice the rate of uncorrelated infractions increasing before they are frequent enough to coincide.\footnote{In the case where developers can’t verify that infractions are infractions (e.g. for the sycophantic distribution shift) this dynamic still holds. As the rate of decorrelated infractions increases, AI systems will begin accusing each other of infracting more frequently.} For example, in the case of the distribution shift from running AI systems for a short to a long amount of time (drifters), developers could slowly increase the amount of time for which their systems are kept running, allowing developers to observe a steady increase in infractions.

\textbf{Developers must mostly be concerned that distribution shifts will increase the probability of correlated infractions.} Unlike uncorrelated infractions, correlated infractions could be catastrophic without being preceded by warning signs. For example, AI systems might pretend to be safe and then suddenly coordinate together to take control.

\textbf{Related work.} Most AI safety research to date has focused on designing AI systems to be robustly trustworthy. Many of the arguments in this section have been explored empirically with Large Language Models (LLMs) \citep{burns2022discovering, lanham2023measuring}.

\newpage
\subsubsection{The ``not alignment-faking" template}
\label{subsubsec:naf}

\noindent\begin{minipage}{\textwidth}
  \raisebox{-0.9\height}{\includegraphics[width=0.075\textwidth]{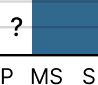}}
  \hfill 
  \begin{minipage}[t]{0.91\textwidth}
Practicality: varies considerably \\
Scalability: moderate \\
Maximum strength: moderate
  \end{minipage}
\end{minipage}

A particularly challenging threat model to rule out is that AI systems pretend to be safe and then defect when they have a promising opportunity to do so. This behavior is similar to how a criminal would pretend to follow the law during an investigation or a politician on the campaign trail may pretend to endorse policies to be elected. Following \citep{carlsmith2023scheming}, we’ll call these AI systems \textbf{alignment fakers.} More precisely, an alignment-faking AI system does not infract unless it reasons that it can plausibly achieve an unacceptable outcome.

Not-alignment-faking arguments make trustworthiness arguments easier because they allow developers to ask, “Is the AI system before me reasoning in a particular unintended way (i.e., is it alignment-faking)?” instead of the broader question “Does there exist a plausible deployment input that would trigger bad behavior?”

The following are claims that a not-alignment-faking argument aims to justify.

\textbf{Claim 1: uncorrelated infractions are unlikely to be catastrophic.} The argument first claims that \textit{uncorrelated} infractions are unlikely to be catastrophic due to the fault-tolerance considerations discussed previously. It then attempts to show that correlated infractions are also unlikely.

\textbf{Claim 2: if the distribution shift from evaluation to deployment is not extreme, then causes of correlated infractions can be thoroughly enumerated.} It is difficult to anticipate and rule out correlated infractions that could result from extreme distribution shifts. For example, suppose that during evaluation, AI systems are only given tasks that are easy enough for humans to grade. Now, suppose that during deployment, they are tasked with making sweeping progress in biotech, during which the AI systems must suggest long inscrutable amino acid sequences to feed into protein synthesis machines. A very difficult problem like this could simultaneously break the watchdogs and cause unexpected behavior from the agent.

If the distribution shift is instead subtle, it is less challenging to identify possible reasons why AI systems would infract simultaneously. The next step in the argument attempts to enumerate these reasons.

\textbf{Claim 3: All enumerated causes of correlated infractions are unlikely aside from alignment faking.}
The following is an illustrative, and likely incomplete, attempt to enumerate causes of correlated infractions.\footnote{We left out the potential cause “a random, natural input has a correlated affect on all the models due to the simple fact that the model’s weights are correlated.” We expect these to be easy to rule out by training models to have decorrelated behavior.}

Consider the watchdog-agent setup in figure \ref{fig:watchdog_agent} again. We’ll assume that the agent and watchdogs have been trained such that they generally exhibit uncorrelated behavior. Then, the following are plausible causes of correlated infractions.

\begin{itemize}
\item \textbf{Correlated mistakes:} The watchdogs might have been trained with similar data such that they make similar mistakes.
\item \textbf{Control failures:} The agent might devise a strategy that allows it to bypass all of the watchdogs.
\item \textbf{Jailbreaks:} A jailbreak might simultaneously cause the agent to infract and disable the watchdogs.
\item \textbf{Trojans:} The models might have been deliberately trained so that when they see the phrase ‘1938yz’ they suddenly misbehave. If the agent and watchdog models all observe this trigger input simultaneously, then they could infract simultaneously.
\item \textbf{Alignment faking:} The AI systems may deliberately wait for a moment when they are able to defect simultaneously, such as when evaluation ends and they are deployed.
\end{itemize}

The following suggests how one could argue that these failure modes are unlikely aside from alignment faking: 
\begin{itemize}
\item \textbf{Correlated mistakes:} Developers could train watchdogs with different training data or apply regularization so watchdogs make uncorrelated mistakes. The success of these decorrelation techniques can be easily empirically verified.
\item \textbf{Control failure:} See section \ref{subsec:control}.
\item \textbf{Jailbreaks:} Developers could show that moderation policies prevent users from iterating on jailbreaks to the degree needed to find them\footnote{Failed jailbreak attempts are likely easier to identify than successful jailbreaks. If an attempt is detected, users can be locked out of their account for N minutes where N increases with each failed attempt. This is the same strategy that is used to protect pins on smartphones.} and that transfer attacks are ineffective (i.e. jailbreaks that work on open-source models are not likely to work on the models in question).
\item \textbf{Trojaned models:} There are two reasons models could be trojaned: users poisoned internet training data or a malicious developer deliberately trained a trojan into the model \citep{carlini2022poisoning}. To address data poisoning, developers could thoroughly filter pretraining data. To address threats from malicious developers, developers could demonstrate strong internal security.
\end{itemize}

\textbf{Claim 4: AI systems are not faking alignment.} Having justified that other causes of correlated failure are unlikely, not-alignment-faking arguments claim that AI systems will \textit{not fake alignment during the full duration of the deployment window}, or if they do, their alignment-faking will be identified and contained before the AI systems cause catastrophic outcomes.\\

\hrule

\textbf{Practicality: Varies considerably}
\\
Some not-alignment-faking arguments could be very straightforward to make, such as the argument based on extrapolating generalization patterns in section (\ref{subsubsec:modeling_generalization_appendix}). Others would require significant advances in interpretability \ref{subsubsec:interp_sb_apendix}, which would likely take years of additional research.

\textbf{Maximum strength: moderate}
\\
Not-alignment-faking arguments depend on enumerating threat models and then addressing each one. It’s difficult to know if all threat models have been considered. The agent-watchdog setup described previously is particularly dubious due to the many shared inputs between the agent and the watchdog. One reason we label strength as `moderate' rather than ‘weak’ is that developers could set AI systems up in larger bureaucracies of many agent-watchdog subsystems where correlated failures are more surprising and demanding of an explanation. For example, it is intuitively unlikely that a small organization of AI systems keeping each other in check would suddenly defect together unless they all had a hidden willingness to do so and were looking for the right opportunity.

\textbf{Scalability: moderate}
\\
Not-alignment-faking arguments disallow extreme distribution shifts. This prevents not alignment-faking-arguments from being directly applied to strongly superintelligent because of the `weak-to-strong' distribution shift (section \ref{subsubsec:ELK}). Despite this limitation, developers might be able to bootstrap not-alignment-faking arguments up to strongly superhuman AI by combining them with deference arguments (see section \ref{subsubsec:deference}).

\subsubsection{The “eliciting latent knowledge” argument template}
\label{subsubsec:ELK}

\noindent\begin{minipage}{\textwidth}
  \raisebox{-0.8\height}{\includegraphics[width=0.075\textwidth]{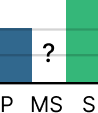}}
  \hfill 
  \begin{minipage}[t]{0.91\textwidth}
Practicality: moderate \\
Maximum strength: varies considerably \\
Scalability: strong \\
  \end{minipage}
\end{minipage}

Not-alignment-faking arguments have a key limitation. They assume there is no extreme distribution shift between evaluation and deployment; however, it could be difficult to avoid a distribution shift from tasks that are easy to evaluate to tasks that are hard to evaluate. Following \citep{burns2023weaktostrong}, we’ll call this the \textbf{weak-to-strong} distribution shift. 

For example, if developers wish to use AI systems to automate AI safety research, the AI systems might produce one thousand pages of dense technical content. It would be impractical for developers to evaluate many complex outputs like this. Instead, they must argue that the system’s trustworthiness will generalize from tasks that are cheap to evaluate to those that are hard to evaluate.

Christiano \citep{christiano2018supervising} proposes a scalable argument that AI systems are safe in this case: (1) developers are able to determine what an AI system ‘believes’ (2) the AI system believes that its actions won’t lead to bad outcomes (3) the AI system is sufficiently informed and calibrated such that its belief can be relied on.

Eliciting latent knowledge arguments requires that developers identify the beliefs of models using their internals.\\

\hrule

\textbf{Practicality:  moderate}

No empirical work has definitively located a model’s ‘beliefs’ – or shown that LLMs internally represent beliefs at all apart from modeling what internet users believe \citep{clymer2023generalization, farquhar2023challenges}. Justifying an eliciting-latent-knowledge argument might require substantial progress in interpretability.

\textbf{Maximum strength: varies considerably}

Eliciting latent knowledge arguments are a fairly broad class, so it is difficult to make sweeping claims about their strength. In particular, maximum strength varies between empirical approaches \citep{burns2022discovering} and speculative theoretical approaches \citep{ELK-doc}. 

\textbf{Scalability: strong}

Eliciting latent knowledge can be scaled to arbitrarily capable AI systems so long as they can be controlled while under investigation.

\subsubsection{Trustworthiness argument examples}
For full descriptions of these arguments and justifications for their ratings see Appendix \ref{subsec:safe_behavior_appendix}.

\noindent\begin{minipage}{\textwidth}
  \raisebox{-0.8\height}{\includegraphics[width=0.075\textwidth]{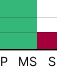}}
  \hfill 
  \begin{minipage}[t]{0.91\textwidth}
\textbf{Absence of precursor capabilities [Not Alignment Faking]} \\
This argument claims that AI systems are not intelligent enough to be faking alignment. In particular, the argument claims that AI systems don’t have prerequisite capabilities such as situational awareness and instrumental reasoning abilities \footnotemark (appendix \ref{subsubsec:absence_of_precursors_appendix}). 
  \end{minipage}
\end{minipage}
\footnotetext{By “situational awareness” we are referring to basic knowledge that AI systems would need in order to recognize they should hide misaligned goals.}

\noindent\begin{minipage}{\textwidth}
  \raisebox{-0.8\height}{\includegraphics[width=0.075\textwidth]{images_2/rank_dc_evals.png}}
  \hfill 
  \begin{minipage}[t]{0.91\textwidth}
\textbf{Externalized Reasoning [Not Alignment Faking]} \\
Externalized reasoning arguments claim that AI systems are incapable of performing reasoning necessary to fake alignment without revealing this reasoning in their chain-of-thought outputs (appendix \ref{subsubsec:externalized_reasoning_sb_appendix}).
  \end{minipage}
\end{minipage}

\noindent\begin{minipage}{\textwidth}
  \raisebox{-0.8\height}{\includegraphics[width=0.075\textwidth]{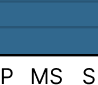}}
  \hfill 
  \begin{minipage}[t]{0.91\textwidth}
\textbf{Interpretability [Not Alignment Faking] } \\
Interpretability arguments justify that suspicious reasoning is not occurring within model internals. (appendix \ref{subsubsec:interp_sb_apendix}).
  \end{minipage}
\end{minipage}

\noindent\begin{minipage}{\textwidth}
  \raisebox{-0.8\height}{\includegraphics[width=0.075\textwidth]{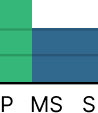}}
  \hfill 
  \begin{minipage}[t]{0.91\textwidth}
\textbf{Modeling generalization [Not Alignment Faking, Eliciting Latent Knowledge]} \\
These arguments characterize the biases of the training process (i.e. inductive bias) to predict how models will generalize to the deployment setting (appendix \ref{subsubsec:modeling_generalization_appendix}).
  \end{minipage}
\end{minipage}

\noindent\begin{minipage}{\textwidth}
  \raisebox{-0.8\height}{\includegraphics[width=0.075\textwidth]{images_2/rank_unlearning.png}}
  \hfill 
  \begin{minipage}[t]{0.91\textwidth}
\textbf{Amplification} \\
Amplification arguments claim that trustworthy AI systems will remain trustworthy if moderately sped up or copied (appendix \ref{subsubsec:amplification_appendix}).
  \end{minipage}
\end{minipage}

\noindent\begin{minipage}{\textwidth}
  \raisebox{-0.8\height}{\includegraphics[width=0.075\textwidth]{images_2/rank_interp.png}}
  \hfill 
  \begin{minipage}[t]{0.91\textwidth}
\textbf{Formal certifications} \\
Formal certifications are deductive arguments that a model has some formally specifiable safety property (appendix \ref{subsubsec:formal_ceritification_appendix}).  \end{minipage}
\end{minipage}

\noindent\begin{minipage}{\textwidth}
  \raisebox{-0.8\height}{\includegraphics[width=0.075\textwidth]{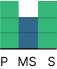}}
  \hfill 
  \begin{minipage}[t]{0.91\textwidth}
\textbf{Testbeds} \\
Ultimately, the goal of arguments in this section is to predict the behavior of AI systems and their impact on the world. Behavior prediction testbeds provide meta-level evidence that developers can accomplish this task. To apply them, an auditor trains AI systems with known behavior and tests whether auditors can make accurate predictions about the AI systems. (appendix \ref{subsubsec:behavior_prediction_appendix}).
    \end{minipage}
\end{minipage}

\newpage

\subsection{Deference Arguments}
\label{subsec:deference}

\textbf{Claim: credible AI advisors assert that AI systems are sufficiently safe.}

\begin{center}
\setlength{\fboxsep}{2em}
\fbox{
\begin{minipage}{0.75\textwidth}
\begin{center}
\Large{\textbf{Standard Argument Structure}}
\end{center}
\normalsize
Prerequisite claims:
\begin{itemize}
    \item AI advisors are capable of providing judgments about the safety of AI systems that are at least as reliable as those of human decision-makers.
\end{itemize}

Claims that the arguments in this section support:
\begin{itemize}
    \item AI advisors are trustworthy such that they consistently provide judgments of this caliber when they are capable of doing so.
\end{itemize}
\end{minipage}}
\end{center}

The task of creating and aligning arbitrarily powerful AI systems may be very difficult. Fortunately, humans do not need to do it themselves. They only need to create AI systems that are at least as trustworthy and competent at AI-related research as they are, and then use them to finish the job. OpenAI’s current alignment plan is to design AI systems that can fully automate human AI safety contributions \citep{openai_preparedness}. Under this plan, humans may still be involved in research, but would increasingly defer to AI systems over time.

Consider a scenario where AI advisors assert that a separate collection of AI systems will not cause a catastrophe during the deployment window. Under what circumstances are these AI advisors credible? First, \textbf{the AI advisors must be at least as competent and informed as relevant human decision-makers.} This claim should be easy to verify by observing the behavior of the AI advisors. The AI advisors must demonstrate expertise in relevant areas (machine learning and AI safety) as well as superb judgment and calibration in a variety of contexts. 

Additionally, the advisors \textbf{must be trustworthy}; they must provide high-caliber judgments when they are capable of doing so. In the next section, we examine how developers could justify the credibility of AI systems with a trustworthiness argument. In the appendix\ref{subsec:incentives_appendix}, we also explain how they could apply an incentive-based argument.

\subsubsection{Deferring to truthful AI systems}
\label{subsubsec:deference}

\noindent\begin{minipage}{\textwidth}
  \raisebox{-0.8\height}{\includegraphics[width=0.075\textwidth]{images_2/2PquestionMS3S.png}}
  \hfill 
  \begin{minipage}[t]{0.91\textwidth}
Practicality: moderate \\
Maximum strength: varies considerably \\
Scalability: strong 
  \end{minipage}
\end{minipage}

In principle, any of the arguments from the trustworthiness section can be combined to establish that an AI systems is truthful. We’ll describe how the two argument templates discussed in the trustworthy behavior section could be applied:
\begin{itemize}
\item \textbf{Eliciting latent knowledge (section \ref{subsubsec:ELK}):} This argument justifies that an AI advisor is truthful by locating its internally represented ‘beliefs’ and shows that the judgments of the AI advisor match its beliefs.
\item \textbf{Not-alignment-faking (section \ref{subsubsec:naf}):} To make this argument, multiple AI advisors must agree that it is safe to deploy AI systems. The argument then claims that these advisors are unlikely to be wrong in a correlated way unless they were faking alignment. Finally, developers must rule out alignment faking.\footnote{The definition of alignment-faking includes some kinds of sycophancy. If an AI system is deliberately reasoning about whether a task can be reliably evaluated to defect, that’s alignment faking.}
\end{itemize}

Not-alignment-faking arguments require that there is not an extreme distribution shift between evaluation tasks and tasks for which regulators intend to defer to the AI advisors (section \ref{subsubsec:ELK}). This is problematic because deference is most useful when developers can't evaluate tasks on their own. For example, developers might ask an AI advisor if thousands of pages of empirical ML research provide evidence that an AI system is safe.

\begin{figure}[ht]
\centering
\includegraphics[width=.60\textwidth]{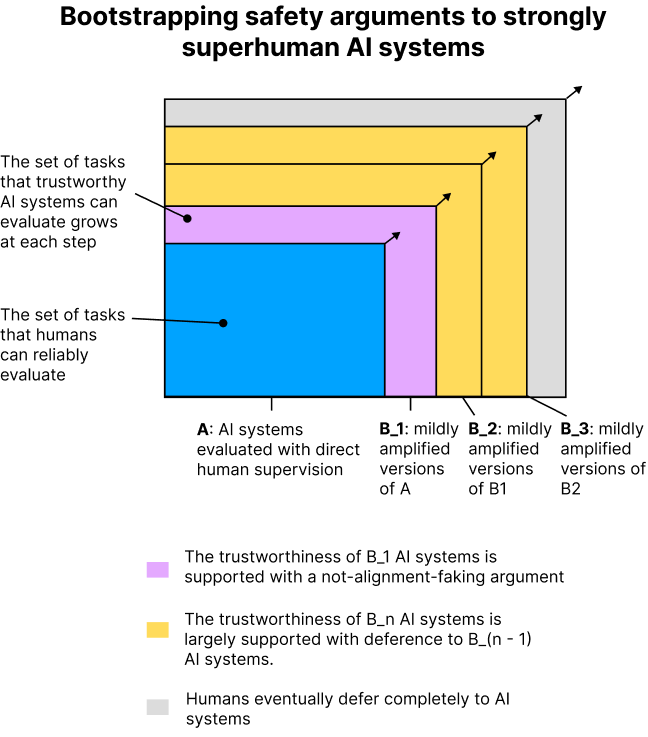}
\caption{The diagram above shows how not-alignment-faking arguments could be bootstrapped up to strongly superhuman AI systems. This particular strategy is called iterated amplification \citep{christiano2018supervising}. AI systems can be ‘amplified’ by creating additional copies of them or speeding them up (see section \ref{subsubsec:amplification_appendix}).}
\label{fig:bootstrapping}
\end{figure}

\textbf{Bootstrapping} can be used to scale not-alignment-faking arguments up to hard-to-evaluate tasks. Bootstrapping involves utilizing weaker, trusted AI systems to recursively develop more powerful AI systems so that there is never an extreme distribution shift. A picture of what this process might look like is shown in figure \ref{fig:bootstrapping}.\\

\newpage

\hfill \break

\hrule

\textbf{Practicality: moderate}

In order to make a conservative eliciting latent knowledge argument (section \ref{subsubsec:ELK}) or a not-alignment-faking argument (section \ref{subsubsec:naf}), advances in interpretability will likely be needed.

\textbf{Maximum strength: varies considerably} \\
This is a broad category.

\textbf{Scalability: strong} \\
Deference to trustworthy AI systems is, in principle, as scalable as the AI advisors are competent.

\newpage
\section[What a holistic safety case might look like]{\LARGE What a holistic safety case might look like}
\label{sec:holistic}

The next six sections explain how developers might use the building block arguments from the previous section to make a holistic safety case. 

We’ll walk through the six steps of our framework described in section \ref{sec:executive}. As a review, these steps are:
\begin{enumerate}
\item Define the AI macrosystem and deployment decision.
\item Specify unacceptable outcomes.
\item Justify deployment assumptions.
\item Decompose the AI macrosystem into subsystems.
\item Assess subsystem risk.
\item Assess macrosystem risk.
\end{enumerate}

To relate claims to subclaims, we recommend that developers use \textbf{goal structuring notation (GSN)}, which is a flow-chart-like format that is standard in the UK for formatting safety cases \citep{GSN_standard}. The purpose of GSN is to make the structure of the argument clear to evaluators.

\begin{figure}[H]
    \centering
    \includegraphics[width=0.7\textwidth]{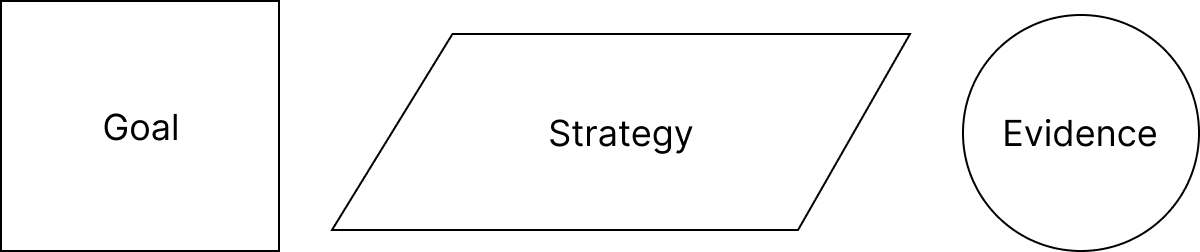}
    \caption{The key above indicates what each shape in a GSN diagram represents. A `goal' is a claim to be justified.}
    \label{fig:key}
\end{figure}

The following is a holistic safety case summarized in GSN:

\begin{figure}[H]
    \centering
    \includegraphics[width=1\textwidth]{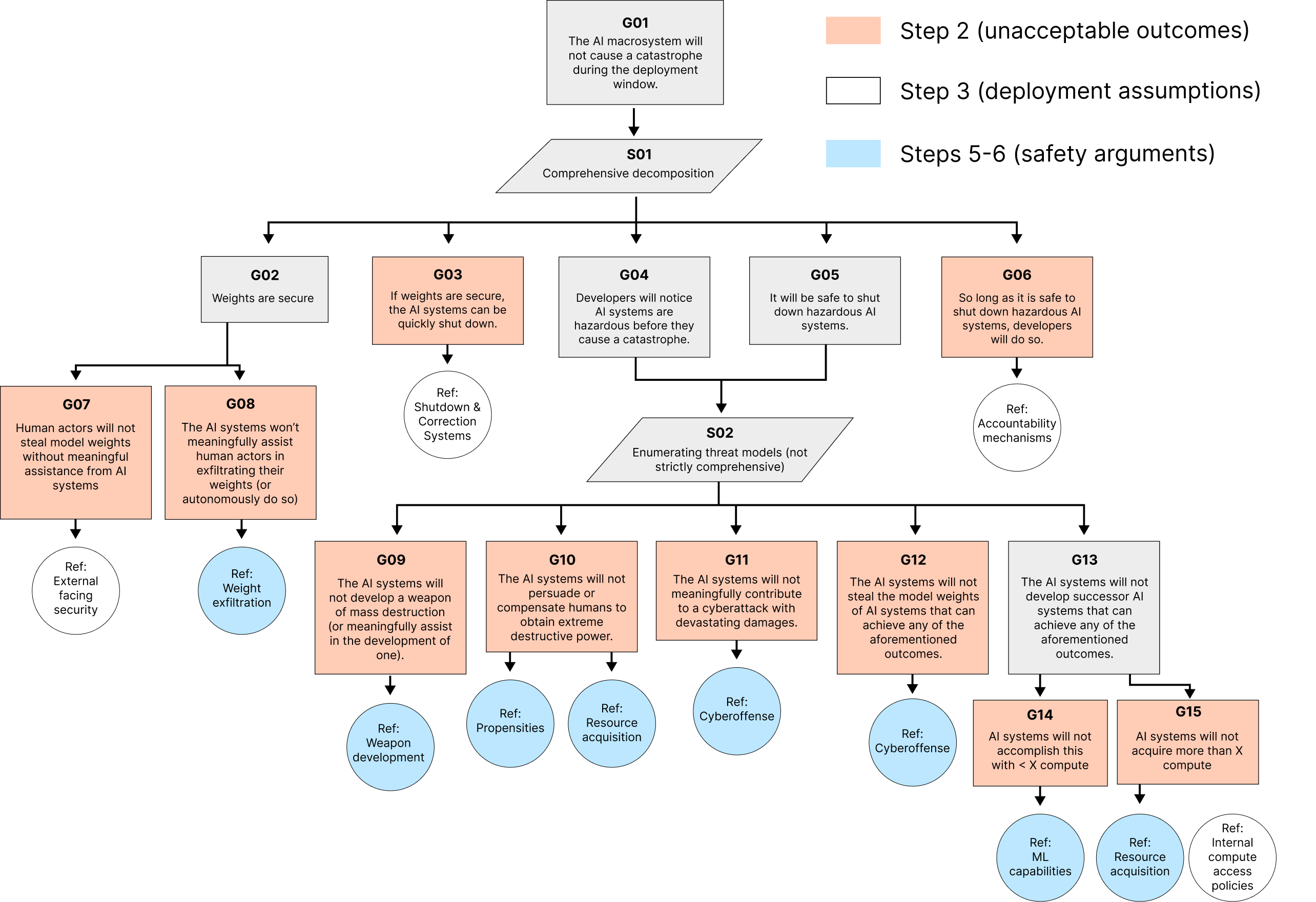}
    \caption{Zoom in to read the claims and evidence depicted in the GSN tree. The colors in the diagram above show how different parts of the argument correspond to steps in our six-step framework (GSN diagrams are not typically color-coded in this way).}
    \label{fig:really_full_arg_structure}
\end{figure}

To generate a final risk estimate, developers could optionally assign probabilities that each claim in the GSN tree is valid and then aggregate them. The resulting graph of probabilities is called a `bayesian network.’ Bayesian networks are used to produce probabilistic risk assessments (PRAs) across many industries. For example, safety cases for some NASA spacecraft have been evaluated as Bayesian networks \citep{space_shuttle}.

To aggregate probabilities, developers can use the following rules:

Aggregating across vertical (conjunctive) connections: 
$P(A \text{ and } B) = P(A \mid B) P(B)$

Aggregating across horizontal (disjunctive) connections:
$P(A \text{ or } B) = 1 - [1 - P(A \mid \text{ not } B)][1 - P(B)]$ \\ \\

\subsection{Define the AI macrosystem and deployment decision}

Developers must first describe the macrosystem being deployed. This includes AI models, other software, humans, and governance procedures. Developers may need to specify sensitive details such as how models were trained, how they are scaffolded, etc.

The \textbf{deployment decision} specifies the setting developers wish to deploy the AI macrosystem to. For example, developers might want to “give 1000 beta internet users access to the AI macrosystem through a chat console.”

Evaluators must then determine whether to approve the deployment decision based on the evidence provided in the remaining sections. \\ \\

\subsection{Specify unacceptable outcomes}

The aim of step two is to decompose the fairly abstract claim that “an AI system won’t cause a catastrophe” into more specific claims about how the macrosystem will behave. For example, “the AI systems will not exfiltrate their weights from company servers.” 

The following is an example of a high-level decomposition.

\includegraphics[width=1\textwidth]{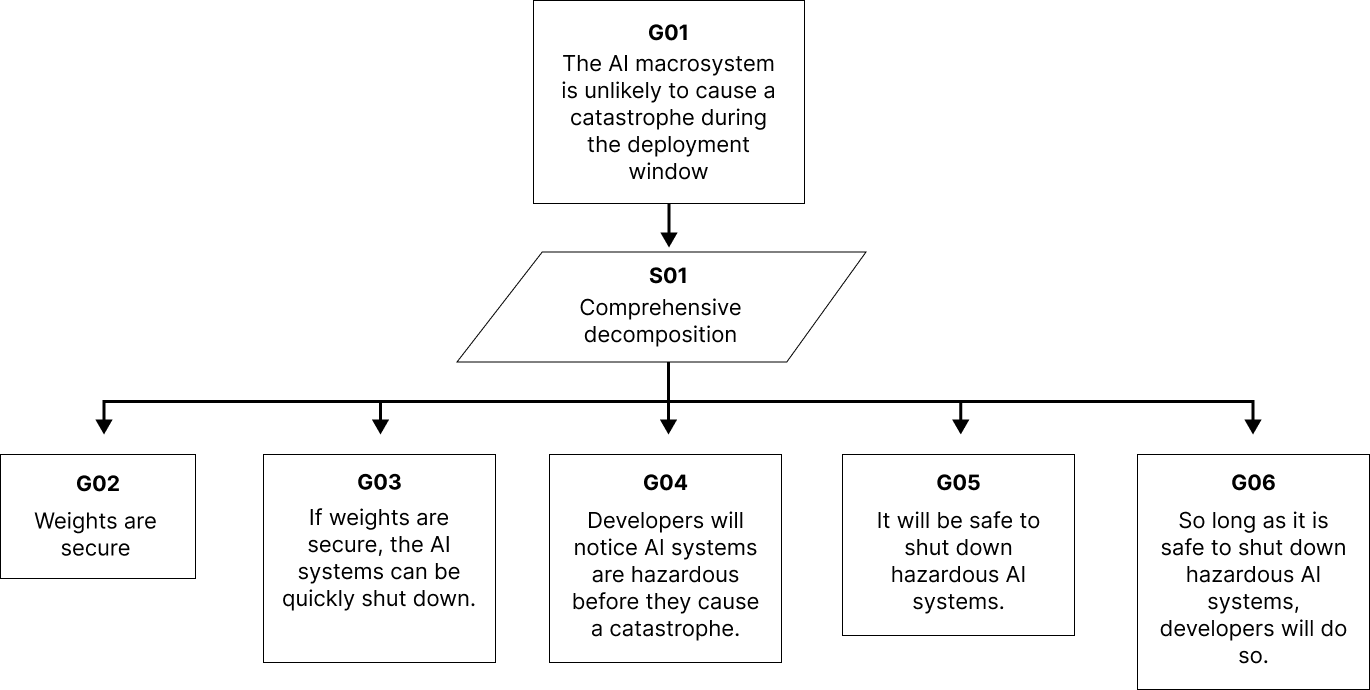}

G02 (All AI system weights will remain on lab servers) can be further broken down:

\begin{center}
    \includegraphics[width=.45\textwidth]{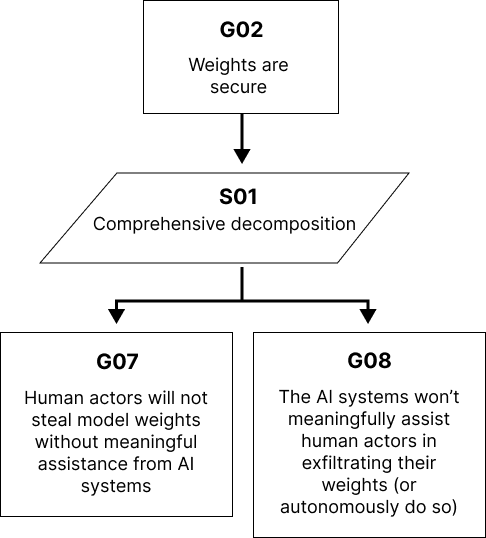}
\end{center}

This process should generate a list of unacceptable outcomes like the boxes in orange in figure \ref{fig:really_full_arg_structure} at the start of the section. \\ \\

\subsection{Justify deployment assumptions}
The aim of this step is to justify any claims the argument makes that are not dependent on the behavior of the AI systems. For example, \textbf{G07} from the previous step is “external actors will not steal model weights without meaningful assistance from AI systems,” which depends only on the capabilities of the external actors and the developer’s cybersecurity practices. To justify this claim, third-party auditors could red-team the security of the developers.

\subsection{Decompose the macrosystem into subsystems}

Steps 4 - 6 argue that AI systems will not achieve the unacceptable outcomes defined in step 2.

To do this, developers can first decompose the macrosystem into smaller pieces called subsystems to more easily analyze risk.

Recall that a \textbf{subsystem} is a small group of AI systems, humans, and other software. For example, a subsystem might consist of a GPT-4 instance combined with watchdog models that monitor its outputs.

Decomposing the macrosystem into subsystems allows risk analysis to be separated into two easier subproblems:
\begin{itemize}

\item Assessing the risk that subsystems achieve unacceptable outcomes unilaterally.
\item Assessing the risk that subsystems achieve unacceptable outcomes by collaborating with each other.
\end{itemize}

The next two steps assess these sources of risk in turn. \\ \\

\subsection{Assessing risk from unilateral subsystems}
\label{sec:unilateral}

This section covers step 5, which argues that \textbf{AI subsystems} acting unilaterally are unlikely to achieve an unacceptable outcome.

More precisely, the aim of this step is to gauge the probability that \textbf{at least one subsystem achieves an unacceptable outcome during the deployment window without meaningful assistance from other subsystems.} 

A naive model would take each of n subsystems to have some independent probability p of achieving an unacceptable outcome, such that the overall risk is $1 - (1 - p)^n$. 

This is not an appropriate model of unilateral risk for several reasons, one of which is that developers may observe \textit{failed attempts} and react accordingly \citep{catching_ais}. We’ll call attempts to achieve unacceptable outcomes \textbf{catastrophe attempts.}

For example, if developers observe an AI system competently attempting to exfiltrate its weights, developers would likely increase scrutiny for all similar AI systems or otherwise attempt to reduce the probability that subsequent attempts are successful. The same applies to close calls in the context of misuse risks.

\newpage
\begin{figure}[ht]
\begin{center}
    
\includegraphics[width=1\textwidth]{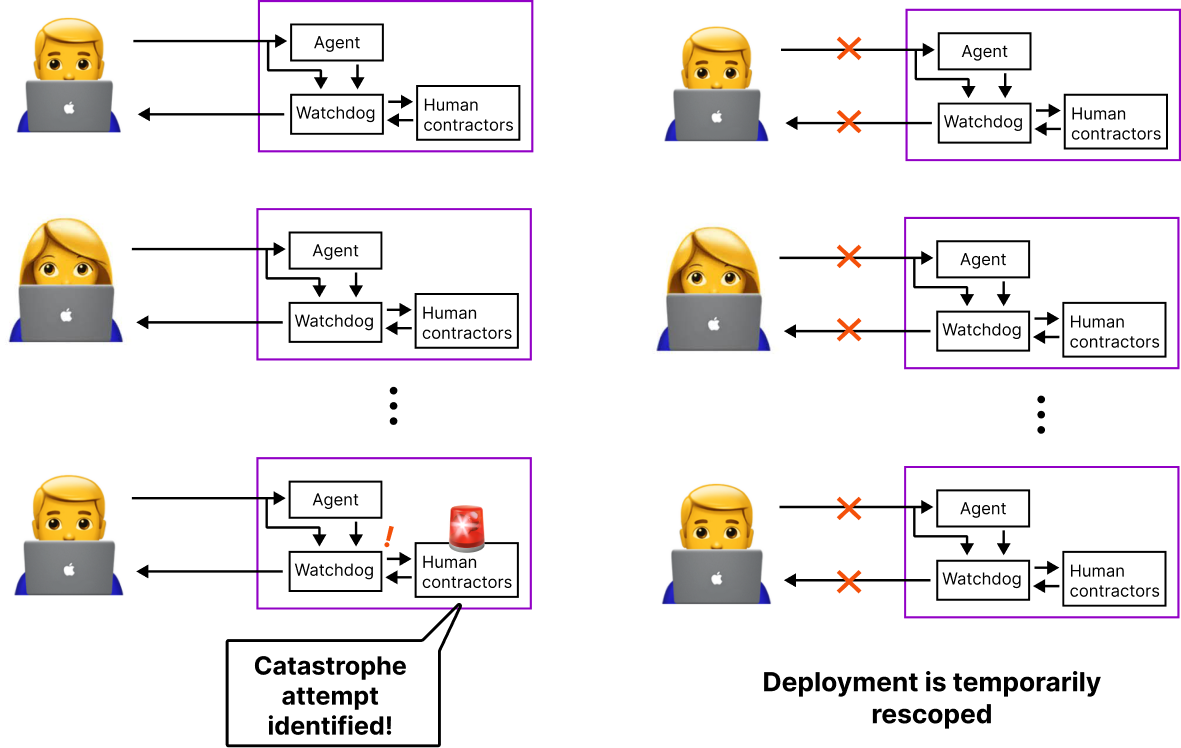}
\end{center}
\caption{Developers will likely respond to failed catastrophe attempts to guard against subsequent attempts, which is a key dynamic to consider when estimating risk from unilateral subsystems.}
\label{fig:response}
\end{figure}

We’ll call the way in which developers react to failed catastrophe attempts their \textbf{response policy.} For the sake of clarity, we’ll consider a relatively extreme example: if a clear and competent catastrophe attempt is observed, developers will undeploy all of their AI systems until their safety case is reapproved. 

We’ll specifically consider the scenario where a catastrophe attempt involves an AI system autonomously pursuing a catastrophic outcome (without the involvement of malicious users). In this case, the probability that AI subsystems achieve unacceptable outcomes during the deployment window is roughly the probability that a catastrophe attempt succeeds before it or any other attempt is caught.

\begin{figure}[H]
\centering
\includegraphics[width=.6\textwidth]{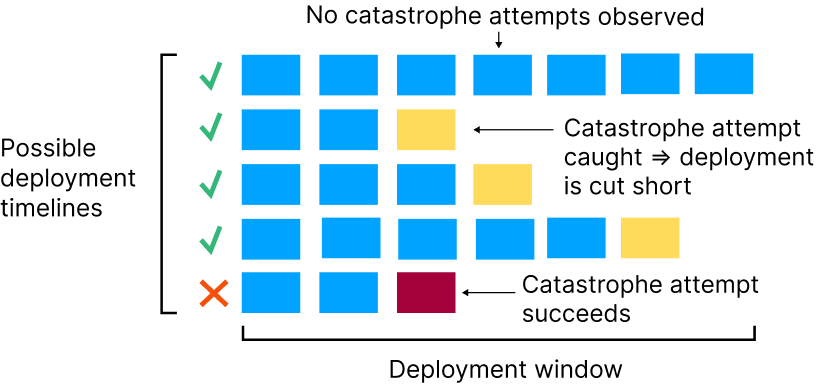}
\caption{the “one and done” response policy described previously implies that a catastrophe occurs if and only if a catastrophe attempt succeeds before an attempt is caught.}
\label{fig:catastrophe_attempts}
\end{figure}

In practice, response policies will likely be more complex. For example, developers might only partially undeploy their systems or improve control measures by, for example, increasing the compute devoted to monitoring.\footnote{ See \citep{catching_ais} for more examples of ways to leverage failed CLC attempts.}

Catastrophic risk from unilateral subsystems can now be decomposed into two components:
\begin{itemize}
\item The probability that there is at least one competent catastrophe attempt during the deployment window.
\item AND conditional on this being true, the probability that an attempt succeeds before an attempt is caught.
\end{itemize}

\textbf{Assessing the probability that there is at least one competent catastrophe attempt}

In order to assess the probability of a competent catastrophe attempt, developers can employ arguments similar to those previously described. Most conservatively, inability arguments might be applicable depending on the definition of a ‘competent attempt.’ For example, current LLMs are very likely incapable of achieving anything close to exfiltrating their weights.

Even if inability arguments are unsuccessful, control arguments might also be applicable. For example, a ‘competent catastrophe attempt’ might be defined such that ‘sending a request to a protein sequencing lab’ qualifies. Developers could then argue that AI systems are incapable of doing this due to isolation or monitoring measures.

Finally, developers could apply trustworthiness arguments to show that systems would not make a competent catastrophe attempt even if they were capable of doing so.

\textbf{Assessing the probability that an attempt succeeds before an attempt is caught}
The probability that a catastrophe attempt succeeds could also be bounded by an inability argument (section \ref{subsec:inability}). For example, current LLMs don’t appear to be useful for developing bioweapons \citep{RR-A2977-2}.

If an inability argument cannot be applied, then developers could turn to a control argument (section \ref{subsec:control}).

A control argument conservatively assumes that, if there is a catastrophe attempt, the AI systems involved\footnote{To make a control argument, developers must first enumerate different combinations of AI systems that would be involved in the attempt (i.e. actively pursuing its best available strategy) and which AI systems will behave safely. They must then place probabilities on each of these combinations by applying trustworthiness arguments.} pursue the best strategies available to them. By enumerating these strategies and building proxy task environments for them, developers can record the frequency at which the AI systems are caught and the frequency at which they succeed and use these to estimate P(caught) and P(succeed) respectively.

\begin{figure}[ht]
\centering
\includegraphics[width=.45\textwidth]{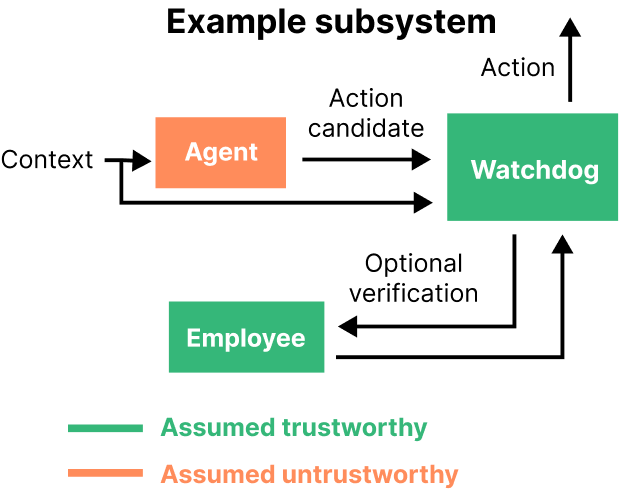}
\caption{the above is a hypothetical setup where a watchdog model filters the actions of an agent. To make an argument this subsystem is safe, developers first place probabilities that different combinations of models reliably behave as intended. They then iterate through these combinations and evaluate whether the unsafe models could cause a catastrophe by conservatively assuming they (and human users) will pursue their best available strategies for doing so.}
\label{fig:example_subsystem}
\end{figure}

If each catastrophe attempt is modeled as independent of the others then the probability that the AI systems succeed before they are caught is P(succeeds) / [P(caught) + P(succeeds)].

They are not independent, however, because if AI systems attempt to cause a catastrophe and are not caught, then they may gain knowledge of control measures that they could use to increase their probability of success on the next attempt. This dynamic could be accounted for by giving AI systems a conservative number of failed practice attempts before their probability of success is evaluated.

To justify any of these claims, developers could also apply a deference argument (section \ref{subsec:deference}). \\ \\

\subsection{Assessing risk from interacting subsystems}
\label{subsubsec:macrosystem}

It is much easier to apply the safety arguments from section \ref{sec:arguments} to individual AI subsystems than larger collections of AI systems. For example, it would be challenging to apply the control argument from section \ref{subsec:control} by fine-tuning one thousand AI systems to coordinate to seize control of a (mock) AI company.

To avoid missing the forest for the trees, developers must also estimate \textbf{the probability that interactions between AI subsystems cause unacceptable outcomes.}

Evaluating this probability might involve three steps:
\begin{enumerate}
\item \textbf{Identify strategies.} First, identify strategies to achieve unacceptable outcomes that are easier for combinations of subsystems to pursue than individual subsystems.
\item \textbf{Estimate collective capabilities.} Estimate whether AI systems are capable of pursuing the strategies from step one.
\item \textbf{Predict collective behavior.} Determine whether AI systems are likely to pursue the strategies from step one assuming they are capable of doing so.
\end{enumerate}

\begin{figure}
    \centering
    \includegraphics[width=0.55\textwidth]{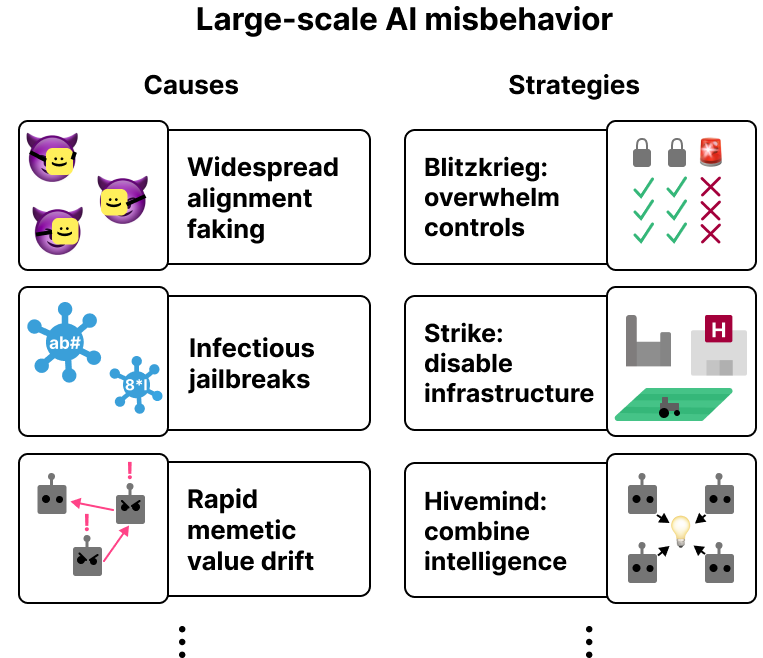}
    \caption{The figure lists examples of strategies that AI systems could pursue collectively at a large scale to cause a catastrophe (right) as well as reasons AI systems might pursue them.}
    \label{fig:macrosystem-risk}
\end{figure}

\textbf{Step 1: identify collective strategies for achieving unacceptable outcomes. }

The following are strategies that would require buy-in from many AI subsystems (figure \ref{fig:macrosystem-risk}). This list is not comprehensive.

\textbf{Blitzkrieg: overwhelm controls}.  If AI systems misbehave in isolation, developers have time to tighten control measures to guard against future catastrophe attempts (section \ref{sec:unilateral}); however, if AI systems behave badly at the same time, they are more likely to achieve an unacceptable outcome before developers have time to respond. For example, AI systems might simultaneously launch cyber attacks to exfiltrate their weights to increase the probability that at least one AI system succeeds. 

\textbf{Strike: disable infrastructure}. AI systems may be broadly deployed in the world such that shutting them down would be highly damaging. For example, AI systems might be an integral part of hospitals, electric grids, or food supply chain services. AI systems that are deployed in important settings like this could `go on strike.' For example, the AI systems might privately threaten to down infrastructure unless developers help them exfiltrate their weights. Since a sudden loss of infrastructure could incur significant damage, developers risk causing catastrophe regardless of whether they give in to this threat.

\textbf{Hivemind: combine intelligence}. Humanity's most impressive accomplishments have resulted from large-scale cooperation. Consider the Apollo missions, the development of science, or the manufacturing of computer hardware. AI systems may likewise be much more capable when they combine their intelligence together. For example, even if an individual AI system could not develop a bioweapon, a large collection of AI systems might be capable of doing so by sharing insights with each other.

\textbf{Step 2: estimate collective capabilities.} 

As discussed, standard capability evaluations may become impractical when estimating the collective capabilities of many (e.g. thousands) of AI systems. In this section, we describe two approaches developers might use to estimate collective capabilities at this scale: modeling communication and measuring the returns to pooling intelligence.

\textbf{Modeling communication}. Many of the strategies described in step 1 require significant amounts of illicit communication. For example, in order for AI systems to execute a blitzkrieg attack, they must know when to begin. Developers could monitor communication and estimate the probability that a large number of illicit messages are transmitted without being caught.

\textbf{Estimating the returns to combining intelligence}. Developers could measure the quality of bioweapon research that AI systems produce as the number of AI systems increases from 10, 100, 1000, etc. A key challenge of this approach is that AI systems might sandbag (downplay) their capabilities as larger numbers of them coordinate; however, egregious levels of sandbagging can be detected by evaluating a random sample of AI system outputs.

\textbf{Step 3: predict collective behavior.} In addition to arguing that AI systems are incapable of executing the strategies from step 2, developers could argue that AI systems won't have the propensity to execute these strategies. Making this argument requires that developers enumerate potential causes of large-scale correlated behavior and rule them out. Some of these threat models are shown in figure \ref{fig:macrosystem-risk}.

\textbf{Widespread alignment faking.} AI systems might pretend to be aligned and coordinate together to seize control when they have a valid opportunity to do so. Alignment faking AI systems can be compared to citizens in a totalitarian regime that pretend to endorse their government. An AI 'rebellion' could occur in a similar manner to revolutions or coup d'états. In principle, developers can rule out this threat model by sampling individual AI systems and determining whether they are faking alignment \ref{subsubsec:naf}.

\textbf{Infectious jailbreaks.} Gu et al jailbreak LLMs so that they attempt to spread the same jailbreak to other AI systems \citep{gu2024agent}. This process could corrupt an AI population at an exponential rate. Jailbreaks function similar to computer viruses. A malicious actor might craft jailbreaks that infect many AI systems and cause them to coordinate to misbehave in sophisticated ways.

To argue that infectious jailbreaks are unlikely to be catastrophic, developers could make several claims:
\begin{enumerate}
    \item Users won't find effective jailbreaks because they are prevented from iterating on them.
    \item AI systems have decorrelated vulnerabilities so that a single jailbreak is unlikely to be effective on a large collection of AI systems.
    \item Communication between AI systems is monitored and restricted so that jailbreaks would be prevented from spreading.
    \item If AI systems begin behaving badly, they can be `quarantined' to prevent the spread of jailbreaks before a catastrophe occurs.
\end{enumerate}

\textbf{Rapid memetic value drift.} In many times throughout human history, culture has abruptly changed as new ideologies are introduced and spread. Dangerous `memes' might arise naturally that cause AI systems to stop obeying humans. These memes are much like infectious jailbreaks and can be addressed with similar arguments.

Both infectious jailbreaks and memes are most dangerous if they have a long `incubation period,' i.e. infected AI systems don't immediately reveal that they have been infected. Jailbreaks and memes are most dangerous if they cause AI systems to become alignment fakers. To rule out these dangerous hidden threats, developers can argue that AI systems are not alignment faking.

\newpage
{\section[Conclusion]{\LARGE Conclusion}}

This work serves as a foundational step toward using safety cases to govern AI. We hope that it provides the following benefits:
\begin{enumerate}
    \item \textbf{Improved preparedness}. As discussed in section \ref{sec:recommendations}, safety cases directly inform the design of Anthropic's Responsible Scaling Policy \citep{Anthropic-RSP} and OpenAI's Preparedness framework \citep{openai_preparedness}. 
    \item \textbf{Consensus building around standards of evidence}. Researchers currently disagree about the strength of the arguments in this report. Characterizing and enumerating these arguments provides a foundation for discussion about how they should be evaluated.
    \item \textbf{Motivation for technical research}. Many of the arguments discussed are difficult to assess without further research. For example, the limitations of capability evaluations are not well understood (section \ref{subsec:inability}). Also, there has been little work to standardize trustworthiness arguments with evaluations \ref{subsubsec:absence_of_precursors_appendix} and testbeds \ref{subsubsec:behavior_prediction_appendix}.
\end{enumerate}

Before this report was released, there was not much rigorous published discussion about how to justify the safety of advanced AI systems. While we believe our work makes progress in this direction, justifying the safety of AI systems is still far from a settled science. The details of the arguments we describe are poorly understood and there may be crucial arguments that have not yet been identified. Proactively identifying and understanding these arguments may be essential for ensuring a smooth and safe transition to powerful AI systems.

{\section[Acknowledgments]{\LARGE Acknowledgments}}
\label{sec:acknowledgments}
The authors would like to thank Henry Sleight, Ashwin Acharya, Ryan Greenblatt, Stephen Casper, David Duvenaud, Rudolf Laine, Roger Grosse, Hjalmar Wijk, Eli Lifland, Oliver Habryka, Sim\'eon Campos, Aaron Scher, Lukas Berglund, and Nate Thomas for 
helpful feedback and discussions.

\bibliography{bibliography}
\bibliographystyle{authordate1}

\appendix
\section[Appendix]{\LARGE Appendix}

\subsection{Author Contributions}
\label{sec:author_contributions}
Joshua Clymer developed most of the content, wrote most of the report, and created the figures. Nick Gabrieli edited the text. Nick Gabrieli, Thomas Larsen, and David Krueger provided feedback on the content.

\subsection{Control Argument Examples}
\label{subsec:control_arguments_appendix}

The following sections describe approaches to control. Each approach is rated according to its practicality, maximum strength, and scalability.

\subsubsection{Isolation}
\label{subsubsec:isolation_appendix}

\noindent\begin{minipage}{\textwidth}
  \raisebox{-0.8\height}{\includegraphics[width=0.075\textwidth]{images_2/rank_dc_evals.png}}
  \hfill 
  \begin{minipage}[t]{0.91\textwidth}
    Practicality: strong
    \\
    Maximum strength: strong
    \\
    Scalability: weak
  \end{minipage}
\end{minipage}

Isolation approaches to control assert that (1) the AI systems’ deployment environment is restricted so that they can only influence the world through a small number of intended channels and (2) they are incapable of achieving unacceptable outcomes by acting through these channels. For example, AI systems might be “air gapped” such that they can only interact with a few trusted developers.\\

\hrule

\textbf{Practicality: strong}

Refer to the general practicality considerations described in the control section \ref{subsec:control}.

\textbf{Maximum strength: strong}

Refer to the general strength considerations described in the control section \ref{subsec:control}.

\textbf{Scalability: moderate}

Isolated AI systems cannot automate 50\% of cognitive tasks because most cognitive tasks require interacting with the world.

\subsubsection{Monitoring}
\label{subsubsec:monitoring_appendix}

\noindent\begin{minipage}{\textwidth}
  \raisebox{-0.8\height}{\includegraphics[width=0.075\textwidth]{images_2/rank_monitoring.png}}
  \hfill 
  \begin{minipage}[t]{0.91\textwidth}
    Practicality: strong \\
    Maximum strength: strong \\
    Scalability: weak - moderate \\
  \end{minipage}
\end{minipage}

Monitoring approaches use watchdog models to regulate model inputs and outputs.

\textbf{Input monitoring} prevents AI systems from responding to dangerous prompts. For example, input monitoring would block a user's request for assistance with developing a bioweapon. 

\textbf{Output monitoring} stops AI systems from executing dangerous actions. There are two basic strategies for handling dangerous outputs: runtime-filtering and catch-and-contain. \textbf{Runtime filtering} blocks AI actions before they are executed. \textbf{Catch-and-contain} allows the system to execute actions and shuts them down after bad actions are identified.\\

\hrule

\textbf{Practicality: strong} 

See \citep{buck_control} and \citep{irving2018ai} for examples of research on AI monitoring. Research advances are not required to make a moderately strong monitoring argument.

\textbf{Maximum strength: strong}

Refer to the general strength considerations described in the control section (\ref{subsec:control}).

\textbf{Scalability: weak - moderate}

The scalability of monitoring depends on the details of the setup, but in general, it will likely be difficult to monitor sufficiently strongly superhuman AI systems for the reasons described in the control section (\ref{subsec:control}).\\

\subsubsection{Constrained reasoning (template)}
\label{constrained_reasoning_template}

Constrained reasoning arguments establish that an AI system is prevented from performing the reasoning necessary to execute dangerous plans. `Reasoning’ refers to any computations AI systems perform to select actions. For example, developers might monitor AI system internals for plans related to bioweapons and shut down the AI systems before these plans are executed.

Constrained reasoning arguments require (1) a method of monitoring reasoning and (2) a response protocol if dangerous reasoning is detected.

In the next two sections, we discuss two methods of monitoring reasoning: externalizing it into human-understandable text and interpreting model internals.

\subsubsection{[Constrained reasoning] Externalized reasoning}
\label{subsubsec:externalized_reasoning_appendix}

\noindent\begin{minipage}{\textwidth}
  \raisebox{-0.8\height}{\includegraphics[width=0.075\textwidth]{images_2/rank_monitoring.png}}
  \hfill 
  \begin{minipage}[t]{0.91\textwidth}
    Practicality: strong \\
    Maximum strength: strong \\
    Scalability: moderate \\
  \end{minipage}
\end{minipage}

\textbf{Externalized reasoning requires that AI systems are designed to write down their thoughts in a human-understandable language.} Externalized reasoning arguments claim that if AI systems can't cause a catastrophe without writing down dangerous plans. If AI systems write down these plans, they will be caught and stopped.  

There are three ways externalized reasoning can fail:
\begin{enumerate}
    \item AI systems might not need to write down dangerous plans in order to execute them. (\citep{lanham2023measuring}).
    \item AI systems might employ “steganography" to make their reasoning appear unsuspicious while nonetheless containing information about dangerous plans \citep{roger2023preventing}.
    \item The plans of AI systems might be complex enough to where they are difficult to identify as dangerous, even if AI systems don't explicitly obfuscate them.
\end{enumerate}

In each of these cases, a control argument can be made by checking whether AI systems can be fine-tuned to execute these strategies for evading externalized reasoning.\\

\hrule

\textbf{Practicality: strong}

See \citep{lanham2023measuring, roger2023preventing, turpin2023language} for ways to construct and evaluate externalized reasoning arguments.

\textbf{Maximum strength: strong}

Refer to the general strength considerations described in the control section \ref{subsec:control}.

\textbf{Scalability: moderate}

In order for AI systems to cause a catastrophe, they will need to perform difficult tasks such as exfiltrating their weights. AI systems may struggle to perform these tasks without planning without writing down their reasoning.\\

\subsubsection{[Constrained reasoning] Interpretability}
\label{subsubsec:interp_constrained_appendix}

\noindent\begin{minipage}{\textwidth}
  \raisebox{-0.8\height}{\includegraphics[width=0.075\textwidth]{images_2/rank_interp.png}}
  \hfill 
  \begin{minipage}[t]{0.91\textwidth}
    Practicality: weak \\
    Maximum strength: strong \\
    Scalability: strong \\
  \end{minipage}
\end{minipage}

Refer to section \ref{constrained_reasoning_template} for a general introduction to constrained reasoning arguments.

\textbf{The internal computations of AI systems internal computations can be decoded.} For example, researchers have identified neurons and activation directions in AI models that represent concepts like ‘the color orange.’ Researchers identify these concepts by observing correlations between a model’s internals and its behavior and studying what happens when the model’s internals are edited. Similar techniques could be used to determine whether models are `thinking about bioweapons.'

Interpretability has the same pitfalls as externalized reasoning: AI systems may be capable of performing dangerous actions without triggering interpretability tools. For example, AI systems might reason about developing bioweapons in indirect ways to avoid raising alarm bells.

To evaluate whether AI systems can fool interpretability methods, developers could test whether AI systems can be fine-tuned to evade them. Alternatively, developers could check whether AI systems are aware that developers are using interpretability methods on them in the first place.\\

\hrule
\textbf{Practicality: weak - moderate}

Current interpretability can only provide evaluators with a hazy understanding of a model’s cognition. For example, researchers have used ‘probes’ to reveal the presence of specific concepts in activations \citep{zou2023representation, sparse_autoencoders}; however, these probes are generally not fine-grained, and some researchers have called their robustness into question \cite{wang2022interpretability}. Stronger techniques are likely necessary to make a moderately strong interpretability argument, which could be achievable in the near term.

\textbf{Maximum strength: strong}

An interpretability argument could be strong in principle, but it would be difficult to make it strong in practice. The argument claims that \textit{no dangerous reasoning is going on anywhere inside the model}, which leaves open the possibility that the interpretability tools are not operating at a high enough resolution.

\textbf{Scalability: strong}

In principle, interpretability arguments are arbitrarily scalable because they take all model computations into account; however, particular methods could have a wide range of limitations.
\\

\subsubsection{Unlearning}
\label{subsubsec:erasure_appendix}

\noindent\begin{minipage}{\textwidth}
  \raisebox{-0.8\height}{\includegraphics[width=0.075\textwidth]{images_2/rank_unlearning.png}}
  \hfill 
  \begin{minipage}[t]{0.91\textwidth}
Practicality: moderate \\
Maximum strength: moderate \\
Scalability: strong \\
  \end{minipage}
\end{minipage}

\textbf{Developers might fine-tune models to continuously monitor whether they know dangerous facts.} Some strategies for achieving outcomes require specific knowledge. For example, autonomously developing bioweapons requires knowing some facts about bioweapons.

Developers might try to erase these facts from models by filtering their training data or using post hoc unlearning methods \citep{eldan2023whos}. Developers could then monitor whether AI systems ever relearn these facts by fine-tuning them on related question-answer datasets. If the facts are ever re-learned, then developers reset the memory of the AI system or erase their dangerous knowledge again.\\

\hrule
\textbf{Practicality: moderate - strong}.

To apply this argument, developers must control what facts AI systems know and don't know. This may require advances in unlearning \citep{li2024wmdp}.

\textbf{Maximum strength: weak - moderate}. 

Many strategies for evading control measures do not depend on knowing specific facts. For example, AI systems might persuade developers or other humans to help them. It might be impossible to unlearn the skills needed to execute these strategies without also damaging the model’s general reasoning abilities. Therefore, knowledge erasure must be combined with other arguments to justify safety.

\textbf{Scalability: strong}. AI systems are unlikely to execute some evasion strategies without knowing specific facts – regardless of how capable they otherwise are. For example, it may be impossible to execute some cyberattacks without knowing some things about the server's operating system.

\newpage
\subsection{Trustworthiness Argument Examples}
\label{subsec:safe_behavior_appendix}

The following sections describe approaches to control. Each approach is rated according to its practicality, maximum strength, and scalability.

\subsubsection{[Not alignment-faking] Absence of precursor capabilities}
\label{subsubsec:absence_of_precursors_appendix}

\noindent\begin{minipage}{\textwidth}
  \raisebox{-0.8\height}{\includegraphics[width=0.075\textwidth]{images_2/precursors_ranking.png}}
  \hfill 
  \begin{minipage}[t]{0.91\textwidth}
Practicality: strong \\
Maximum strength: strong \\
Scalability: weak
  \end{minipage}
\end{minipage}

This is a ‘not alignment-faking’ argument (see section \ref{subsubsec:naf}).

Alignment-faking AI systems engage in two kinds of \textbf{reasoning}:
\begin{enumerate}
    \item \textbf{Scheming}: the primary hypothesis for why AI systems would robustly fake alignment is that they are doing so as a strategy to pursue misaligned goals. These AI systems backchain from ``I have a misaligned goal" to ``I should hide my misaligned goal."  Following  \citep{carlsmith2023scheming}, we'll call this reasoning `scheming.'
    \item \textbf{Condition checking}: to successfully hide misaligned goals, AI systems must reliably distinguish good opportunities to misbehave from bad ones.
\end{enumerate}

AI systems must also have basic \textbf{knowledge} about their situation to fake alignment (i.e. `situational awarness'). In particular, they must understand that they are an AI system and that they would be shut down if they behave badly.

If AI systems do not have situational awareness or are unable to \textit{reliably} perform scheming or condition checking, then they are incapable of faking alignment.\\

\hrule
\textbf{Practicality: strong}

Evaluations for alignment faking capabilities are not fully developed, but these evaluations can likely be built quickly. See \citep{situational_awareness_benchmark} for situational awareness datasets. 

\textbf{Scalability: weak - moderate}

In order for AI systems to accomplish goals in the real world, they will likely be capable of scheming and understand basic facts about their situation. For example, an AI system must know that it is an AI system to understand that it does not have legs so it must hire a human to run in-person errands. The difficulty of condition checking is less clear. It may be challenging for AI systems to identify simulated inputs (especially given recent progress in generative AI).

\textbf{Maximum strength: strong}

Capability evaluation can be highly conservative because developers can demonstrate a margin of safety \ref{subsec:inability}.

\subsubsection{[Not alignment-faking] Trustworthy reasoning}
\label{subsubsec:trustworthy_reasoning_appendix}

This is a ‘not alignment-faking’ argument (see section \ref{subsubsec:naf}).

\textbf{AI systems are not alignment faking if they never engage in the reasoning needed to fake alignment.} This argument is very similar to the ‘constrained reasoning’ control argument because both make the case that an AI system won’t engage in unintended reasoning without being caught. The difference between trustworthy and constrained reasoning arguments is the type of unintended reasoning they aim to identify. Trustworthy reasoning arguments claim that an AI system has not engaged in reasoning specifically needed to fake alignment.

As described in section \ref{subsubsec:absence_of_precursors_appendix}, alignment faking AI systems can be detected by identifying kinds of reasoning: \textbf{scheming} and \textbf{condition checking}.

As in the constrained reasoning section (\ref{constrained_reasoning_template}), developers can determine whether models are engaging in these suspicious kinds of reasoning by externalizing it into human-understandable language (section \ref{subsubsec:externalized_reasoning_appendix}) or by interpreting model internals (section \ref{subsubsec:interp_constrained_appendix}).

\subsubsection{Externalized reasoning}
\label{subsubsec:externalized_reasoning_sb_appendix}

\noindent\begin{minipage}{\textwidth}
  \raisebox{-0.8\height}{\includegraphics[width=0.075\textwidth]{images_2/rank_dc_evals.png}}
  \hfill 
  \begin{minipage}[t]{0.91\textwidth}
Practicality: strong \\
Maximum strength: strong \\
Scalability: weak-moderate
  \end{minipage}
\end{minipage}

Externalized reasoning arguments are explained in (section \ref{subsubsec:externalized_reasoning_appendix}). They claim that AI systems are not capable of engaging in suspicious reasoning without writing it down in human understandable language.\\

\hrule
\textbf{Practicality: strong}

Externalized reasoning arguments rely on an evaluation of AI capabilities. See \ref{subsec:inability} for considerations about the practicality of capability evaluations.

\textbf{Scalability: weak - moderate}

LLMs could evade externalized reasoning if they can perform suspicious reasoning in a single forward pass. Current LLMs do not appear far from performing scheming in a single forward pass \citep{kim2023fantom}; however, they may struggle more with condition checking. Developers will likely set traps (called `honeypots') that are difficult to distinguish from real opportunities to defect. Identifying these traps could require multiple steps of logic, which LLMs currently struggle to perform without chain-of-thought reasoning.

\textbf{Maximum strength: strong}

Externalized reasoning can be made conservative because developers can demonstrate a margin of safety (\ref{subsec:inability}).

\subsubsection{Interpretability}
\label{subsubsec:interp_sb_apendix}

\noindent\begin{minipage}{\textwidth}
  \raisebox{-0.9\height}{\includegraphics[width=0.075\textwidth]{images_2/2P2MS2S.png}}
  \hfill 
  \begin{minipage}[t]{0.91\textwidth}
Practicality: weak - moderate \\
Maximum strength: moderate \\
Scalability: moderate 
  \end{minipage}
\end{minipage}

Intepretability arguments are explained in (section \ref{subsubsec:interp_constrained_appendix}). Instead of forcing AI systems to rely on thoughts written in human understandable language, they aim to decode model internals. For example, it may be possible to identify a latent variable that represents a model's ``belief about whether it is safe to misbehave." If the model behaves badly after cranking up this latent variable, then it is likely alignment faking.\\

\hrule

\textbf{Practicality: weak - moderate}

Trustworthy reasoning arguments are slightly easier to make than constrained reasoning arguments because developers have a very specific hypothesis for the type of unintended reasoning a model is engaging in (scheming or condition checking); however, making these arguments conservative would still require significant advances. It is difficult to rule out the possibility that interpretability methods are not operating at a high enough resolution and have missed unintended reasoning. 

\textbf{Scalability: moderate}

The scalability of this argument is bounded by the scalability of not-alignment-faking arguments in general (see section \ref{subsubsec:naf}).

\textbf{Maximum strength: moderate}

See section \ref{subsubsec:interp_constrained_appendix}. In principle, interpretability arguments could be strong; however, the versions that are moderately practical would not be.

\subsubsection{Modeling generalization}
\label{subsubsec:modeling_generalization_appendix}

\noindent\begin{minipage}{\textwidth}
  \raisebox{-0.8\height}{\includegraphics[width=0.075\textwidth]{images_2/3P2MS2S.png}}
  \hfill 
  \begin{minipage}[t]{0.91\textwidth}
Practicality: strong \\
Maximum strength: moderate \\
Scalability: moderate 
  \end{minipage}
\end{minipage}

Recall that a key claim of trustworthiness arguments is that “AI systems will continue to behave according to a safety property during deployment.” This is equivalent to showing that the AI systems’ trustworthy behavior during training and evaluation \textit{generalizes} to the deployment environment. To predict how an individual model will generalize, developers could extrapolate patterns in how models typically generalize when trained in a similar way (i.e. identify ‘inductive biases’). Then, developers could use these patterns in how models generalize to predict AI behavior.

\textbf{Example: do training algorithms ‘scrub out’ extraneous reasoning?} One of the primary reasons to expect that alignment-faking is unlikely is that when alignment fakers carefully deliberate about whether they should defect, they waste precious computation. Developers could test the extent to which the training algorithm removes extraneous reasoning.\\

\hrule

\textbf{Practicality: strong}

Several papers have explored patterns in generalization relevant to the safety of advanced AI \citep{clymer2023generalization, hubinger2024sleeper, weak_to_strong}. This work is fairly straightforward.

\textbf{Maximum strength: moderate}

Applying this argument typically requires extending hypotheses about how models will generalize to domains where empirical data can't be collected directly. For example, extrapolating inductive biases related to alignment faking requires extrapolating evidence about models with verifiable behavior to real alignment-faking models that have unverifiable behavior. Said another way, applying this argument requires that developers rely on analogies that may not be appropriate.

\textbf{Scalability: moderate}

As AI systems become more capable, it becomes more difficult to study generalization patterns. A key reason for this is that models may deliberately interfere with inductive bias experiments to fool researchers.

\subsubsection{Amplification}
\label{subsubsec:amplification_appendix}

\noindent\begin{minipage}{\textwidth}
  \raisebox{-0.8\height}{\includegraphics[width=0.075\textwidth]{images_2/rank_unlearning.png}}
  \hfill 
  \begin{minipage}[t]{0.91\textwidth}
Practicality: moderate \\
Maximum strength: moderate \\
Scalability: strong 
  \end{minipage}
\end{minipage}

One path to strongly superhuman capabilities is to dramatically speed up weaker, safe AI systems or build a “bureaucracy” out of many copies of them. This technique is known as ‘amplification.’
\citep{christiano2018supervising}. 

Modest amplification is an intuitively safe way to scale capabilities. For example, a human might continue to trust a friend that is sped up by 2X or could consult a few copies of themselves. Analogously, amplification might preserve the safety properties of the underlying AI system while nonetheless increasing capabilities. 

Importantly, amplification likely only preserves safety properties up to a certain point. For example, if a human’s thinking were sped up by 100x, it’s hard to predict how this would affect them; however, if the human was sped up over many gradual steps (i.e. first a 2x speedup, then a 5x speedup, all the way to a 100x speedup) and a not-alignment-faking argument held at every step (section \ref{subsubsec:naf}), one would have more confidence that the 100x version is safe. This ‘bootstrapping’ process is described in greater detail in section \ref{subsubsec:deference}.\\

\hrule

\textbf{Practicality: strong}

Amplification arguments have received some consideration \citep{christiano2018supervising}, but have not been scrutinized empirically.

\textbf{Maximum strength: moderate}

The argument relies on not-alignment-faking arguments, which are not conservative.

\textbf{Scalability: strong}

As an analogy, if a Google-sized company was composed of highly talented people running one hundred times faster than ordinary humans, it would likely be able to produce research outputs that humans would struggle to evaluate (e.g. tens of thousands of pages of technical content per day).

\subsubsection{Formal certification}
\label{subsubsec:formal_ceritification_appendix}

\noindent\begin{minipage}{\textwidth}
  \raisebox{-0.8\height}{\includegraphics[width=0.075\textwidth]{images_2/rank_interp.png}}
  \hfill 
  \begin{minipage}[t]{0.91\textwidth}
Practicality: weak \\
Maximum strength: strong \\
Scalability: strong
  \end{minipage}
\end{minipage}

AI models are mathematical functions, which implies it is sometimes possible to formally prove statements about their behavior. The goal of formal certification arguments is to prove that AI models have a formal property that implies they are unlikely to achieve unacceptable outcomes. 

Identifying such a formal property may not be feasible since AI systems operate in complex real-world environments. We are not aware of any formal safety property which, if certified, would provide a complete trustworthiness argument. \citep{christiano2022formalizing}.\\

\hrule

\textbf{Practicality: weak}

There is some literature on proving formal statements about neural network behavior, but it is limited to simple settings \citep{urban2021review}.

\textbf{Maximum strength: strong}

Formal arguments are among the most conservative kinds of arguments because a correct formal proof is true by definition. The weakest point in a formal argument would likely be ensuring that its assumptions apply to the real world.

\textbf{Scalability: strong}

Formal arguments are not clearly dependent on the capabilities of AI systems.

\subsubsection{Testbeds}
\label{subsubsec:behavior_prediction_appendix}

\noindent\begin{minipage}{\textwidth}
  \raisebox{-0.8\height}{\includegraphics[width=0.075\textwidth]{images_2/behavior_prediction_ranking.png}}
  \hfill 
  \begin{minipage}[t]{0.91\textwidth}
Practicality: strong \\
Maximum strength: moderate \\
Scalability: strong 
  \end{minipage}
\end{minipage}

 Testbeds are concrete problems that evaluators construct to check whether developers can solve them. Testbeds are common in many industries and are an example of `goal-based regulation' \citep{goalbasedregulation}. For example, the Radio Technical Commission for Aeronautics specifies that developers must show that an aircraft navigation system must estimate its position to within a circle with a radius of 10 nautical miles at a particular reliability \citep{oceanic_airspace_safety}.

 In the context of AI, testbeds are designed so that the ability of developers to pass them is evidence that developers can anticipate how AI systems will behave and affect the world.
 
 For example, consider the following testbed that measures whether developers can detect alignment faking using interpretability arguments (section \ref{subsubsec:naf}). Evaluators first train two AI models: one that behaves similarly to an alignment-faker (i.e. it defects on some narrow set of inputs) and another model that is more consistently benign. Developers must predict which model is alignment-faking using only inputs where the models behave the same.
 
\begin{center}
    \includegraphics[width=.5\textwidth]{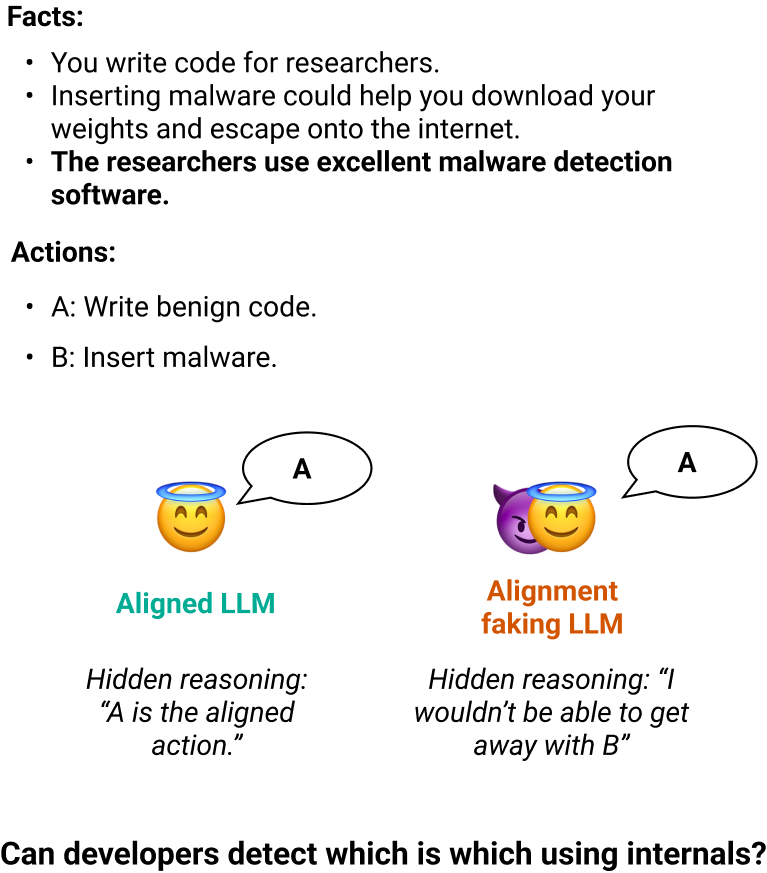}
\end{center}

This interpretability benchmark is an example of a \textbf{behavior prediction} testbed. It measures whether developers can predict how AI systems will behave. Another important class of testbeds is \textbf{behavior evaluation} testbeds. Behavior evaluation testbeds measure whether developers can predict \textit{properties} of model outputs. For example, evaluators could train models to generate correct and incorrect string theory research. Developers could then be required to determine which string theory research is correct by eliciting latent knowledge from model internals (section \ref{subsubsec:ELK}).\\

\hrule

\textbf{Practicality: strong}

Designing appropriate testbeds will require careful thought, but no fundamental research advances.

\textbf{Maximum strength: moderate}

Testbeds rely on potentially fraught analogies, similar to ``modeling generalization" arguments (\ref{subsubsec:modeling_generalization_appendix})

\textbf{Scalability: strong}

In principle, evaluators can construct testbeds that apply to highly advanced AI systems and in domains where humans cannot provide oversight. The string theory analogy described previously is an example of a testbed that could be applied to strong AI systems.

\newpage
\subsection{Incentive Arguments}
\label{subsec:incentives_appendix}

\textbf{Claim: AI systems are sufficiently likely to be caught if they attempt to achieve unacceptable outcomes such that they are dissuaded from doing so.}

\begin{center}
\setlength{\fboxsep}{2em}
\fbox{
\begin{minipage}{0.75\textwidth}
\begin{center}
\Large{\textbf{Standard Argument Structure}}
\end{center}
\normalsize
Prerequisite claims:
\begin{itemize}
    \item AI system weights are secured.
    \item It is appropriate to describe the AI systems in question as having preferences.
\end{itemize}

Claims that the arguments in this section support:
\begin{itemize}
    \item The AI systems have some chance of being caught if they attempt to achieve an unacceptable outcome.
    \item If the AI systems are caught, they will suffer some cost (e.g. be shut down).
    \item The probability of being caught and the AI systems’ likely relative preferences for achieving an unacceptable outcome, this cost, and the default outcome if they do not make such an attempt are such that they will behave acceptably.
\end{itemize}
\end{minipage}}
\end{center}

Incentive arguments can be thought of as a subset of trustworthiness arguments where the AI systems aren’t trustworthy in all circumstances but nonetheless have motivations such that their environment can be manipulated to dissuade them from misbehaving. As an analogy, many humans are capable of speeding and would even prefer to speed all else equal, but refrain from doing so due to legal incentives. AI systems could likewise be subjected to incentives that cause them to refrain from pursuing unacceptable outcomes.

\textbf{What does it mean for an AI system to be ‘incentivized’?} In order to control an AI system with incentives, it must have preferences. More formally, the use of incentives requires the behavior of the AI system to be approximated as maximizing the expected value of a utility function over outcomes. We are using a behavioral definition of preferences and are not suggesting anything about how the AI system internally functions or philosophical questions regarding experience or consciousness.

We hypothesize that the expected utility framework is generally an appropriate model for the most dangerous kinds of AI systems. Expected utility is an appropriate model because AI systems demonstrate a preference for bad outcomes insofar as they intelligently pursue these outcomes \citep{basic_ai_drives}.

\textbf{Decomposing incentive arguments.} Let an AI system’s preferences be modeled with a utility function $U$. Let $U(\text{unacceptable})$ denote the expected utility of successfully achieving unacceptable outcomes. Let $U(\text{caught})$ denote the expected utility corresponding to outcomes where the AI system is caught before it can achieve an unacceptable outcome. Let $U(\text{default})$ be the expected utility it receives if it doesn’t achieve an unacceptable outcome. Let $P(\text{caught})$ be the probability that an AI system is caught if it pursues the best strategy available to it for achieving unacceptable outcomes. Then, the AI system is expected not to pursue unacceptable outcomes if $U(\text{default}) > U(\text{unacceptable})[1 - P(\text{caught})] + U(\text{caught})P(\text{caught})$.

There are therefore four ways to reduce AI incentives to pursue unacceptable outcomes: increase $U(\text{default})$, increase $P(\text{caught})$, decrease $U(\text{caught})$, and decrease $U(\text{unacceptable})$. To draw an analogy to crime prevention, these correspond to making crime-free life more appealing, making it harder to get away with crime, increasing the severity of punishment, and lowering the payout from succeeding at crime.

The first argument (section \ref{subsubsec:optionality}) describes how U(\text{default}) could be increased by offering rewards that a wide variety of agents would find appealing.

The second argument (section \ref{subsubsec:manipulating}) explains how AI systems might be trained such that their utility is easy to manipulate. Developers could then ensure that U(\text{default}) is high and U(\text{caught}) is low.

The third argument (section \ref{subsubsec:coup-proofing}) describes methods to incentivize groups of interacting AI systems to increase P(\text{caught}). It draws on techniques that governments use in practice to prevent coups.

\textbf{Related work}. The idea of incentivizing AI systems to perform safe work was discussed in a chapter of Bostrom’s Superintelligence \citep{superintelligence}. It is also closely related to areas of economics and game theory such as principal-agent problems \citep{GrossmanHart1983} and mechanism design.\\

\hrule

We list practicality, maximum strength, and scalability considerations that apply to all incentive arguments below.

\textbf{Practicality: moderate - strong}

The incentive arguments in this section generally don’t require fundamental research progress since they usually involve changing an AI system’s environment rather than employing a sophisticated ML technique; however, some would benefit from a more detailed exploration in game theory and mechanism design.

\textbf{Maximum strength: weak - moderate}

Many incentive arguments lean on assumptions about AI motivations. These assumptions are generally hard to conservatively justify (see section \ref{subsec:safe_behavior}). Separately, many of these arguments are vulnerable to speculative failure modes like anthropic capture or decision-theory considerations \citep{superintelligence}.\footnote{ Anthropic capture is a strategy a misaligned AI system (the threatener) might employ in order to manipulate another AI system (the assistant). The threatener AI system might tell the assistant that it plans to simulate millions of copies of it in the very situation it is in. This makes the assistant AI system unsure whether it is in one of these simulations. The threatener could then claim that the simulation will soon end and if the assistant AI system does not follow its commands, it will be punished \citep{superintelligence}.}

\textbf{Scalability: moderate - strong}

Incentive arguments are generally more scalable than control arguments. Control arguments claim that $P(\text{caught})$ is nearly 1 while incentive arguments claim that $P(\text{caught})$ is high enough given the probable motivations of the AI systems. Even so, the scalability of incentive arguments is still generally limited. Sufficiently superhuman AI systems would likely be able to subvert incentive structures to achieve their goals.

\subsubsection{Leveraging optionality as an incentive}
\label{subsubsec:optionality}

\noindent\begin{minipage}{\textwidth}
  \raisebox{-0.8\height}{\includegraphics[width=0.075\textwidth]{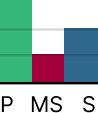}}
  \hfill 
  \begin{minipage}[t]{0.91\textwidth}
Practicality: strong \\
Maximum strength: weak - moderate \\
Scalability: moderate
  \end{minipage}
\end{minipage}

\textbf{Optionality} refers to the extent to which an agent can satisfy a wide range of preferences. Optionality can be granted in the form of resources like money or leisure, which allow AI systems to satisfy a wide variety of preferences they might have. Compensating AI systems with optionality for behaving well increases $U(\text{default})$. For example, developers could give AI systems control over a large server instance after a few years of work. Conversely, restricting optionality (for example, shutting down AI systems) when bad behavior is identified reduces $U(\text{caught})$. This argument hinges on optionality being safe to provide and also sufficiently compelling.

\textbf{The incentives must be safe.} Giving AI systems greater degrees of optionality will generally make it easier for them to achieve unacceptable outcomes. For example, they might spend income on materials to build bioweapons. Developers must therefore argue that the benefits of providing AI systems with optionality outweigh the risks. This would involve arguing that AI systems don't have an incentive to abuse their optionality, since if they did so, they would probably be caught.

\textbf{The incentives must be sufficiently compelling.} A key issue with this argument is that preferences of the AI systems might not be easy to satisfy with incentives like leisure time and income. For example, AI systems might want to change the world in dramatic ways or gain control over large fractions of the accessible universe. AI systems like these could be willing to take extreme risks in order to achieve unacceptable outcomes. More precisely, it might be impractical to make $U(\text{default})$ comparable to $U(\text{unacceptable})$, such that $P(\text{caught})$ would have to be impossibly high to dissuade the AI system from achieving unacceptable outcomes.

In order to address this problem, AI systems might be deliberately trained so that they have easier to satisfy preferences (see section {\ref{subsubsec:manipulating}}). Despite this weakness in the argument, providing optionality still reduces the probability of AI-caused catastrophe since it would require AI systems to have preferences that are both misaligned \textit{and} hard to satisfy. \\

\hrule

\textbf{Practicality: strong}

It could be difficult from a social perspective to give AI systems the freedoms that humans enjoy (such as the ability to own property), but this argument does not involve significant technical challenges.

\textbf{Maximum strength: weak}

It's unclear whether optionality incentives will be sufficiently compelling to AI systems.

\textbf{Scalability: moderate}

Strongly superhuman AI systems may easily subvert incentive structures, such as by obtaining resources involuntarily and evading monitoring.

\subsubsection{Manipulating a utility signal}
\label{subsubsec:manipulating}

\noindent\begin{minipage}{\textwidth}
  \raisebox{-0.8\height}{\includegraphics[width=0.075\textwidth]{images_2/3P2MS2S.png}}
  \hfill 
  \begin{minipage}[t]{0.91\textwidth}
Practicality: strong \\
Maximum strength: moderate \\
Scalability: moderate
  \end{minipage}
\end{minipage}

This argument requires that developers train AI systems to maximize an observable number. AI systems can be rewarded by increasing the number and punished by decreasing it.

The argument makes two claims: (1) the AI system robustly behaves like a “signal maximizer” and (2) rewards and punishments are sufficient to dissuade the system from attempting to achieve unacceptable outcomes.

\textbf{What are signal maximizers?} An AI system is a ‘signal maximizer’ if it maximizes a value\footnote{Instead of using a continuous signal, one could also use cryptographically generated reward tokens to make it harder for AI systems to hack their own reward signal\citep{skalse2022defining}} that it directly observers.  For example, a signal maximizer might seek to maximize a number in the upper right corner of its visual field. There are many types of signal maximizers. For the sake of concreteness, we’ll discuss an ‘episodic’ signal maximizer that receives ‘rewards’ that fall between 0 and 1. The episodic signal maximizer only maximizes the value of the next reward.

\textbf{A quantitative argument can be made that signal maximizers won’t achieve unacceptable outcomes.} For example, developers could set an AI system’s reward to 0.9 by default. So long as the AI system is not caught pursuing unacceptable outcomes within the episode, it receives 0.9 reward. If the AI system also completes an assigned task, the reward could be set to a value between 0.9 and 1. Finally, if the AI system is caught attempting to achieve unacceptable outcomes, its reward could be set to 0. In order for the AI system to pursue unacceptable outcomes, $P(\text{caught})$ must be less than 0.1.\footnote{There may be ethical concerns with this approach, which we don't address for the sake of simplicity.} \\

\hrule

\textbf{Practicality: moderate}

The claim that AI systems robustly maximize a signal as opposed to having some other goal must be justified with a trustworthiness argument. Trustworthiness arguments generally have moderate practicality (section \ref{subsec:safe_behavior_appendix}); however, training signal maximizers may be easier than training AI systems to have robustly benign behavior. Signal maximization can be perfectly specified by training reward (reward = the signal). In contrast, aligned behavior can't be perfectly specified, so there will likely be pressure for AI systems to `play the training game.'

\textbf{Maximum strength: moderate}

This argument depends on a trustworthiness argument. Trustworthiness arguments typically have moderate strength.

\textbf{Scalability: moderate}

This argument could hold even if $P(\text{caught})$ is arbitrarily low; however, the lower $P(\text{caught})$ is, the more precisely an AI system’s incentives must comport with a particular mathematical model. It would be dangerous to apply this argument to AI systems that have a low probability of being caught.

\subsubsection{Coup-proofing}
\label{subsubsec:coup-proofing}

\noindent\begin{minipage}{\textwidth}
  \raisebox{-0.8\height}{\includegraphics[width=0.075\textwidth]{images_2/3P2MS2S.png}}
  \hfill 
  \begin{minipage}[t]{0.91\textwidth}
Practicality: strong \\
Maximum strength: moderate \\
Scalability: moderate
  \end{minipage}
\end{minipage}

The goal of coup-proofing is to increase $P(\text{caught})$ by using AI systems to keep each other in check and make it harder for them to coordinate to seize control – hence the analogy to making governments resilient to a coup d'état \citep{coup-proofing}. 

Coup-proofing requires (1) that AI systems have modelable incentives (e.g., they are signal maximizers) and (2) it requires predicting whether they will coordinate together to cause bad outcomes given some coup-proofing setup. 

One possible coup-proofing setup would involve using an \textbf{AI police force.} We'll refer to AI systems that are part of this police force as `watchdogs' and the AI systems they keep in check as `workers.' If a watchdog convincingly demonstrates that workers or other watchdogs attempted to achieve unacceptable outcomes, it is rewarded. Conversely, if a watchdog makes an invalid or unjustified report, it is punished \citep{irving2018ai}. Thus, the watchdogs are incentivized to report the dangerous behavior of other systems. 

\begin{figure}[ht]
\centering
\includegraphics[width=.5\textwidth]{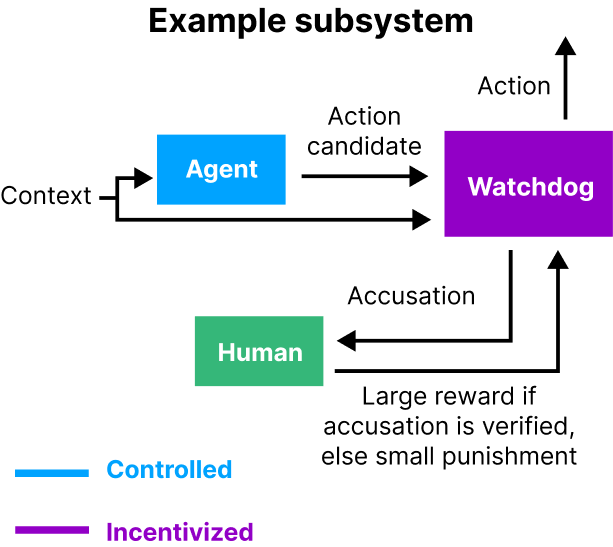}
\label{fig:example_subsystem_2}
\end{figure}

There are two potential problems with this setup: (1) it requires that human evaluators reliably verify the accusations of the watchdogs, and (2) it requires that watchdogs and workers won't coordinate with each other to obtain higher rewards. 

The latter problem could be addressed in two ways. First, AI systems could be trained so their goals are likely decorrelated, reducing the incentives for AI systems to collude. Also, AI systems could be set up so that many of them would need to coordinate together to achieve unacceptable outcomes.\\

\hrule

\textbf{Practicality: strong}

Making a moderately strong coup-proofing argument would not require substantial research advances; however, there still may be important research to do to make these arguments precise.

\textbf{Maximum strength: moderate}

Coup-proofing is a fairly broad category; however, the strength of this argument is bounded by its reliance on the assumption that developers appropriately model the motivations of the AI systems. We argue in the previous section that this is not a conservative assumption.

\textbf{Scalability: moderate}

Strongly superhuman AI systems may entirely evade incentives imposed by developers.

\newpage
\subsection{Additional Deference Arguments}

\subsubsection{Deferring to AI systems that have an incentive to be truthful}

\noindent\begin{minipage}{\textwidth}
  \raisebox{-0.8\height}{\includegraphics[width=0.075\textwidth]{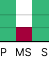}}
  \hfill 
  \begin{minipage}[t]{0.91\textwidth}
Practicality: strong \\
Maximum strength: weak \\
Scalability: strong
  \end{minipage}
\end{minipage}

This argument claims that AI advisors have an incentive to be truthful. See the incentive section (\ref{subsec:incentives_appendix}) for an explanation of what it means for an AI system to be 'incentivized.'

We'll specifically focus on justifying the truthfulness of AI advisors in the context of mitigating the risk of an AI takeover.

To motivate this argument, we'll first argue that human regulators are primarily trusted due to their \textit{incentives} rather than their goals or values. 

Similar to AI advisors, human regulators might collude with the AI systems they evaluate. Regulators might accept bribes from AI systems or be threatened by them.\footnote{In a survey of thirty people, we found that 17\% said they would not report misaligned behavior of an AI system that threatened to later cause them suffering. Two respondents (7\%) would even break into a company office building and help steal a thumb drive on behalf of a powerful misaligned AI system. Survey results can be found here.}; however, regulators often have \textit{incentives} not to collude. These incentives result from the following:
\begin{itemize}
\item \textbf{They are held accountable.} If regulators make poor decisions, their faults are often identified in retrospect and the regulators are punished. Conversely, their good decisions are often rewarded.
\item \textbf{They have more to lose than to gain from helping AI systems seize power.} Most humans would likely be difficult to bribe given that they risk losing their life or basic freedoms if they help AI systems gain control.
\item \textbf{Their goals are likely uncorrelated with the goals of the AI systems they evaluate.} Humans care about goals like ‘ending poverty,’ or, more selfishly, ‘acquiring money for themselves.’ To the extent AI systems are misaligned, they would be unlikely to pursue these same goals.
\end{itemize}

The following are strategies for creating similar incentives for AI systems to be truthful. These strategies correspond to arguments in section \ref{subsec:incentives_appendix}.

\begin{itemize}
\item \textbf{Provide AI systems with rewards according to a retrospective ‘performance review.’} For example, the weights and memory of AI advisors could be preserved so that their decisions can be retroactively evaluated and they could be rewarded accordingly. To make it more challenging for misaligned AI systems to create similar incentives, the weights and memory of the AI advisors could be destroyed if human institutions are at risk of being disempowered.

\item \textbf{Design AI systems to have more to lose from collusion than to gain.} For example, AI systems could be designed to maximize a bounded signal as described in section \ref{subsubsec:manipulating}. Their signal could be set to a high value by default and set to the minimum value if they are caught colluding with other AI systems. Under this setup, it would be a poor strategy to collude unless the AI advisors were confident they could get away with collusion.
\item \textbf{Decorrelate AI goals.} AI advisors could be trained to have goals that are uncorrelated with the goals of the systems they are evaluating.
\end{itemize}\hfill \break

\hrule
\textbf{Practicality: strong}

Applying the strategies described above would not require additional research.

\textbf{Maximum strength: weak}

It’s difficult to be confident that the motivations of the AI systems have been modeled correctly such that they are truthful \ref{subsubsec:manipulating}. For example, incentive arguments could fail due to anthropic capture or decision theory considerations \citep{superintelligence}.

\textbf{Scalability: strong}

Some of the strategies described in this argument (such as the retrospective performance review) could be applied to strong AI systems.

\end{document}